\documentstyle[11pt]{article}

\newcommand{\stln}{\setlength{\unitlength}{2.2ex}}
\newcommand{\fr}{\framebox(1,1){}}
\newcommand{\sfr}{\framebox(1,1){\begin{picture}(1,1)
  \put(0,0){\line(1,1){1}}\end{picture}}}

\newcommand{\onebox}
{\stln \lower1.4ex\hbox{
\begin{picture}(1.6,1.6)
\put(.3,.3){\fr}
\end{picture}}}

\newcommand{\twobox}
{\stln \lower1.4ex\hbox{
\begin{picture}(2.6,1.6)
\put(.3,.3){\fr}
\put(1.3,.3){\fr}
\end{picture}}}

\newcommand{\threebox}
{\stln \lower1.4ex\hbox{
\begin{picture}(3.6,1.6)
\multiput(.3,.3)(1,0){3}{\fr}
\end{picture}}}

\newcommand{\fourbox}
{\stln \lower1.4ex\hbox{
\begin{picture}(4.6,1.6)
\multiput(.3,.3)(1,0){4}{\fr}
\end{picture}}}

\newcommand{\fivebox}
{\stln \lower1.4ex\hbox{
\begin{picture}(5.6,1.6)
\multiput(.3,.3)(1,0){5}{\fr}
\end{picture}}}

\newcommand{\sixbox}
{\stln \lower1.4ex\hbox{
\begin{picture}(6.6,1.6)
\multiput(.3,.3)(1,0){6}{\fr}
\end{picture}}}

\newcommand{\genrowbox}
{\stln \lower1.4ex\hbox{
\begin{picture}(7.6,1.6)
\multiput(.3,.3)(1,0){3}{\fr}
\put(3.3,.3){\framebox(3,1){$\cdots$}}
\put(6.3,.3){\fr}
\end{picture}}}

\newcommand{\oneonebox}
{\stln \lower2.6ex\hbox{
\begin{picture}(1.6,2.6)
\put(.3,.3){\fr}
\put(.3,1.3){\fr}
\end{picture}}}

\newcommand{\twoonebox}
{\stln \lower2.6ex\hbox{
\begin{picture}(2.6,2.6)
\put(.3,1.3){\fr}
\put(1.3,1.3){\fr}
\put(0.3,0.3){\fr}
\end{picture}}}

\newcommand{\threeonebox}
{\stln \lower2.6ex\hbox{
\begin{picture}(3.6,2.6)
\multiput(.3,1.3)(1,0){3}{\fr}
\put(.3,.3){\fr}
\end{picture}}}

\newcommand{\fouronebox}
{\stln \lower2.6ex\hbox{
\begin{picture}(4.6,2.6)
\multiput(.3,1.3)(1,0){4}{\fr}
\put(.3,.3){\fr}
\end{picture}}}

\newcommand{\fiveonebox}
{\stln \lower2.6ex\hbox{
\begin{picture}(5.6,2.6)
\multiput(.3,1.3)(1,0){5}{\fr}
\put(.3,.3){\fr}
\end{picture}}}

\newcommand{\sixonebox}
{\stln \lower2.6ex\hbox{
\begin{picture}(6.6,2.6)
\multiput(.3,1.3)(1,0){6}{\fr}
\put(.3,.3){\fr}
\end{picture}}}

\newcommand{\twotwobox}
{\stln \lower2.6ex\hbox{
\begin{picture}(2.6,2.6)
\put(.3,.3){\fr}
\put(.3,1.3){\fr}
\put(1.3,.3){\fr}
\put(1.3,1.3){\fr}
\end{picture}}}

\newcommand{\threetwobox}
{\stln \lower2.6ex\hbox{
\begin{picture}(3.6,2.6)
\multiput(.3,1.3)(1,0){3}{\fr}
\put(.3,.3){\fr}
\put(1.3,.3){\fr}
\end{picture}}}

\newcommand{\fourtwobox}
{\stln \lower2.6ex\hbox{
\begin{picture}(4.6,2.6)
\multiput(.3,1.3)(1,0){4}{\fr}
\put(.3,.3){\fr}
\put(1.3,.3){\fr}
\end{picture}}}

\newcommand{\fivetwobox}
{\stln \lower2.6ex\hbox{
\begin{picture}(5.6,2.6)
\multiput(.3,1.3)(1,0){5}{\fr}
\put(.3,.3){\fr}
\put(1.3,.3){\fr}
\end{picture}}}

\newcommand{\sixtwobox}
{\stln \lower2.6ex\hbox{
\begin{picture}(6.6,2.6)
\multiput(.3,1.3)(1,0){6}{\fr}
\put(.3,.3){\fr}
\put(1.3,.3){\fr}
\end{picture}}}

\newcommand{\threethreebox}
{\stln \lower2.6ex\hbox{
\begin{picture}(3.6,2.6)
\multiput(.3,1.3)(1,0){3}{\fr}
\multiput(.3,.3)(1,0){3}{\fr}
\end{picture}}}

\newcommand{\fourthreebox}
{\stln \lower2.6ex\hbox{
\begin{picture}(4.6,2.6)
\multiput(.3,1.3)(1,0){4}{\fr}
\multiput(.3,.3)(1,0){3}{\fr}
\end{picture}}}

\newcommand{\fivethreebox}
{\stln \lower2.6ex\hbox{
\begin{picture}(5.6,2.6)
\multiput(.3,1.3)(1,0){5}{\fr}
\multiput(.3,.3)(1,0){3}{\fr}
\end{picture}}}

\newcommand{\sixthreebox}
{\stln \lower2.6ex\hbox{
\begin{picture}(6.6,2.6)
\multiput(.3,1.3)(1,0){6}{\fr}
\multiput(.3,.3)(1,0){3}{\fr}
\end{picture}}}

\newcommand{\fourfourbox}
{\stln \lower2.6ex\hbox{
\begin{picture}(4.6,2.6)
\multiput(.3,1.3)(1,0){4}{\fr}
\multiput(.3,.3)(1,0){4}{\fr}
\end{picture}}}

\newcommand{\fivefourbox}
{\stln \lower2.6ex\hbox{
\begin{picture}(5.6,2.6)
\multiput(.3,1.3)(1,0){5}{\fr}
\multiput(.3,.3)(1,0){4}{\fr}
\end{picture}}}

\newcommand{\sixfourbox}
{\stln \lower2.6ex\hbox{
\begin{picture}(6.6,2.6)
\multiput(.3,1.3)(1,0){6}{\fr}
\multiput(.3,.3)(1,0){4}{\fr}
\end{picture}}}

\newcommand{\fivefivebox}
{\stln \lower2.6ex\hbox{
\begin{picture}(5.6,2.6)
\multiput(.3,1.3)(1,0){5}{\fr}
\multiput(.3,.3)(1,0){5}{\fr}
\end{picture}}}

\newcommand{\sixfivebox}
{\stln \lower2.6ex\hbox{
\begin{picture}(6.6,2.6)
\multiput(.3,1.3)(1,0){6}{\fr}
\multiput(.3,.3)(1,0){5}{\fr}
\end{picture}}}

\newcommand{\sixsixbox}
{\stln \lower2.6ex\hbox{
\begin{picture}(6.6,2.6)
\multiput(.3,1.3)(1,0){6}{\fr}
\multiput(.3,.3)(1,0){6}{\fr}
\end{picture}}}

\newcommand{\oneoneonebox}
{\stln \lower3.8ex\hbox{
\begin{picture}(1.6,3.6)
\multiput(.3,.3)(0,1){3}{\fr}
\end{picture}}}

\newcommand{\twooneonebox}
{\stln \lower3.8ex\hbox{
\begin{picture}(2.6,3.6)
\multiput(.3,.3)(0,1){3}{\fr}
\put(1.3,2.3){\fr}
\end{picture}}}

\newcommand{\twotwoonebox}
{\stln \lower3.8ex\hbox{
\begin{picture}(2.6,3.6)
\multiput(.3,.3)(0,1){3}{\fr}
\put(1.3,1.3){\fr}
\put(1.3,2.3){\fr}
\end{picture}}}

\newcommand{\twotwotwobox}
{\stln \lower3.8ex\hbox{
\begin{picture}(2.6,3.6)
\multiput(.3,.3)(0,1){3}{\fr}
\multiput(1.3,.3)(0,1){3}{\fr}
\end{picture}}}

\newcommand{\threeoneonebox}
{\stln \lower3.8ex\hbox{
\begin{picture}(3.6,3.6)
\multiput(.3,.3)(0,1){3}{\fr}
\put(1.3,2.3){\fr}
\put(2.3,2.3){\fr}
\end{picture}}}

\newcommand{\threetwoonebox}
{\stln \lower3.8ex\hbox{
\begin{picture}(3.6,3.6)
\multiput(.3,.3)(0,1){3}{\fr}
\put(1.3,2.3){\fr}
\put(2.3,2.3){\fr}
\put(1.3,1.3){\fr}
\end{picture}}}

\newcommand{\threetwotwobox}
{\stln \lower3.8ex\hbox{
\begin{picture}(3.6,3.6)
\multiput(.3,.3)(0,1){3}{\fr}
\multiput(1.3,.3)(0,1){3}{\fr}
\put(2.3,2.3){\fr}
\end{picture}}}

\newcommand{\threethreeonebox}
{\stln \lower3.8ex\hbox{
\begin{picture}(3.6,3.6)
\multiput(.3,2.3)(1,0){3}{\fr}
\multiput(.3,1.3)(1,0){3}{\fr}
\put(.3,.3){\fr}
\end{picture}}}

\newcommand{\threethreetwobox}
{\stln \lower3.8ex\hbox{
\begin{picture}(3.6,3.6)
\multiput(.3,2.3)(1,0){3}{\fr}
\multiput(.3,1.3)(1,0){3}{\fr}
\put(.3,.3){\fr}
\put(1.3,.3){\fr}
\end{picture}}}

\newcommand{\threethreethreebox}
{\stln \lower3.8ex\hbox{
\begin{picture}(3.6,3.6)
\multiput(.3,2.3)(1,0){3}{\fr}
\multiput(.3,1.3)(1,0){3}{\fr}
\multiput(.3,.3)(1,0){3}{\fr}
\end{picture}}}

\newcommand{\gencolbox}
{\stln \lower8.6ex\hbox{
\begin{picture}(1.6,7.6)
\multiput(.3,4.3)(0,1){3}{\fr}
\put(.3,1.3){\framebox(1,3){$\vdots$}}
\put(.3,.3){\fr}
\end{picture}}}

\newcommand{\sonebox}
{\stln \lower1.4ex\hbox{
\begin{picture}(1.6,1.6)
\put(.3,.3){\sfr}
\end{picture}}}

\newcommand{\stwobox}
{\stln \lower1.4ex\hbox{
\begin{picture}(2.6,1.6)
\put(.3,.3){\sfr}
\put(1.3,.3){\sfr}
\end{picture}}}

\newcommand{\sthreebox}
{\stln \lower1.4ex\hbox{
\begin{picture}(3.6,1.6)
\multiput(.3,.3)(1,0){3}{\sfr}
\end{picture}}}

\newcommand{\sfourbox}
{\stln \lower1.4ex\hbox{
\begin{picture}(4.6,1.6)
\multiput(.3,.3)(1,0){4}{\sfr}
\end{picture}}}

\newcommand{\sfivebox}
{\stln \lower1.4ex\hbox{
\begin{picture}(5.6,1.6)
\multiput(.3,.3)(1,0){5}{\sfr}
\end{picture}}}

\newcommand{\ssixbox}
{\stln \lower1.4ex\hbox{
\begin{picture}(6.6,1.6)
\multiput(.3,.3)(1,0){6}{\sfr}
\end{picture}}}

\newcommand{\sgenrowbox}
{\stln \lower1.4ex\hbox{
\begin{picture}(7.6,1.6)
\multiput(.3,.3)(1,0){3}{\sfr}
\put(3.3,.3){\framebox(3,1){$\cdots$}}
\put(6.3,.3){\sfr}
\end{picture}}}

\newcommand{\soneonebox}
{\stln \lower2.6ex\hbox{
\begin{picture}(1.6,2.6)
\put(.3,.3){\sfr}
\put(.3,1.3){\sfr}
\end{picture}}}

\newcommand{\stwoonebox}
{\stln \lower2.6ex\hbox{
\begin{picture}(2.6,2.6)
\put(.3,1.3){\sfr}
\put(1.3,1.3){\sfr}
\put(0.3,0.3){\sfr}
\end{picture}}}

\newcommand{\sthreeonebox}
{\stln \lower2.6ex\hbox{
\begin{picture}(3.6,2.6)
\multiput(.3,1.3)(1,0){3}{\sfr}
\put(.3,.3){\sfr}
\end{picture}}}

\newcommand{\sfouronebox}
{\stln \lower2.6ex\hbox{
\begin{picture}(4.6,2.6)
\multiput(.3,1.3)(1,0){4}{\sfr}
\put(.3,.3){\sfr}
\end{picture}}}

\newcommand{\sfiveonebox}
{\stln \lower2.6ex\hbox{
\begin{picture}(5.6,2.6)
\multiput(.3,1.3)(1,0){5}{\sfr}
\put(.3,.3){\sfr}
\end{picture}}}

\newcommand{\ssixonebox}
{\stln \lower2.6ex\hbox{
\begin{picture}(6.6,2.6)
\multiput(.3,1.3)(1,0){6}{\sfr}
\put(.3,.3){\sfr}
\end{picture}}}

\newcommand{\stwotwobox}
{\stln \lower2.6ex\hbox{
\begin{picture}(2.6,2.6)
\put(.3,.3){\sfr}
\put(.3,1.3){\sfr}
\put(1.3,.3){\sfr}
\put(1.3,1.3){\sfr}
\end{picture}}}

\newcommand{\sthreetwobox}
{\stln \lower2.6ex\hbox{
\begin{picture}(3.6,2.6)
\multiput(.3,1.3)(1,0){3}{\sfr}
\put(.3,.3){\sfr}
\put(1.3,.3){\sfr}
\end{picture}}}

\newcommand{\sfourtwobox}
{\stln \lower2.6ex\hbox{
\begin{picture}(4.6,2.6)
\multiput(.3,1.3)(1,0){4}{\sfr}
\put(.3,.3){\sfr}
\put(1.3,.3){\sfr}
\end{picture}}}

\newcommand{\sfivetwobox}
{\stln \lower2.6ex\hbox{
\begin{picture}(5.6,2.6)
\multiput(.3,1.3)(1,0){5}{\sfr}
\put(.3,.3){\sfr}
\put(1.3,.3){\sfr}
\end{picture}}}

\newcommand{\ssixtwobox}
{\stln \lower2.6ex\hbox{
\begin{picture}(6.6,2.6)
\multiput(.3,1.3)(1,0){6}{\sfr}
\put(.3,.3){\sfr}
\put(1.3,.3){\sfr}
\end{picture}}}

\newcommand{\sthreethreebox}
{\stln \lower2.6ex\hbox{
\begin{picture}(3.6,2.6)
\multiput(.3,1.3)(1,0){3}{\sfr}
\multiput(.3,.3)(1,0){3}{\sfr}
\end{picture}}}

\newcommand{\sfourthreebox}
{\stln \lower2.6ex\hbox{
\begin{picture}(4.6,2.6)
\multiput(.3,1.3)(1,0){4}{\sfr}
\multiput(.3,.3)(1,0){3}{\sfr}
\end{picture}}}

\newcommand{\sfivethreebox}
{\stln \lower2.6ex\hbox{
\begin{picture}(5.6,2.6)
\multiput(.3,1.3)(1,0){5}{\sfr}
\multiput(.3,.3)(1,0){3}{\sfr}
\end{picture}}}

\newcommand{\ssixthreebox}
{\stln \lower2.6ex\hbox{
\begin{picture}(6.6,2.6)
\multiput(.3,1.3)(1,0){6}{\sfr}
\multiput(.3,.3)(1,0){3}{\sfr}
\end{picture}}}

\newcommand{\sfourfourbox}
{\stln \lower2.6ex\hbox{
\begin{picture}(4.6,2.6)
\multiput(.3,1.3)(1,0){4}{\sfr}
\multiput(.3,.3)(1,0){4}{\sfr}
\end{picture}}}

\newcommand{\sfivefourbox}
{\stln \lower2.6ex\hbox{
\begin{picture}(5.6,2.6)
\multiput(.3,1.3)(1,0){5}{\sfr}
\multiput(.3,.3)(1,0){4}{\sfr}
\end{picture}}}

\newcommand{\ssixfourbox}
{\stln \lower2.6ex\hbox{
\begin{picture}(6.6,2.6)
\multiput(.3,1.3)(1,0){6}{\sfr}
\multiput(.3,.3)(1,0){4}{\sfr}
\end{picture}}}

\newcommand{\sfivefivebox}
{\stln \lower2.6ex\hbox{
\begin{picture}(5.6,2.6)
\multiput(.3,1.3)(1,0){5}{\sfr}
\multiput(.3,.3)(1,0){5}{\sfr}
\end{picture}}}

\newcommand{\ssixfivebox}
{\stln \lower2.6ex\hbox{
\begin{picture}(6.6,2.6)
\multiput(.3,1.3)(1,0){6}{\sfr}
\multiput(.3,.3)(1,0){5}{\sfr}
\end{picture}}}

\newcommand{\ssixsixbox}
{\stln \lower2.6ex\hbox{
\begin{picture}(6.6,2.6)
\multiput(.3,1.3)(1,0){6}{\sfr}
\multiput(.3,.3)(1,0){6}{\sfr}
\end{picture}}}

\newcommand{\soneoneonebox}
{\stln \lower3.8ex\hbox{
\begin{picture}(1.6,3.6)
\multiput(.3,.3)(0,1){3}{\sfr}
\end{picture}}}

\newcommand{\stwooneonebox}
{\stln \lower3.8ex\hbox{
\begin{picture}(2.6,3.6)
\multiput(.3,.3)(0,1){3}{\sfr}
\put(1.3,2.3){\sfr}
\end{picture}}}

\newcommand{\stwotwoonebox}
{\stln \lower3.8ex\hbox{
\begin{picture}(2.6,3.6)
\multiput(.3,.3)(0,1){3}{\sfr}
\put(1.3,1.3){\sfr}
\put(1.3,2.3){\sfr}
\end{picture}}}

\newcommand{\stwotwotwobox}
{\stln \lower3.8ex\hbox{
\begin{picture}(2.6,3.6)
\multiput(.3,.3)(0,1){3}{\sfr}
\multiput(1.3,.3)(0,1){3}{\sfr}
\end{picture}}}

\newcommand{\sthreeoneonebox}
{\stln \lower3.8ex\hbox{
\begin{picture}(3.6,3.6)
\multiput(.3,.3)(0,1){3}{\sfr}
\put(1.3,2.3){\sfr}
\put(2.3,2.3){\sfr}
\end{picture}}}

\newcommand{\sthreetwoonebox}
{\stln \lower3.8ex\hbox{
\begin{picture}(3.6,3.6)
\multiput(.3,.3)(0,1){3}{\sfr}
\put(1.3,2.3){\sfr}
\put(2.3,2.3){\sfr}
\put(1.3,1.3){\sfr}
\end{picture}}}

\newcommand{\sthreetwotwobox}
{\stln \lower3.8ex\hbox{
\begin{picture}(3.6,3.6)
\multiput(.3,.3)(0,1){3}{\sfr}
\multiput(1.3,.3)(0,1){3}{\sfr}
\put(2.3,2.3){\sfr}
\end{picture}}}

\newcommand{\sthreethreeonebox}
{\stln \lower3.8ex\hbox{
\begin{picture}(3.6,3.6)
\multiput(.3,2.3)(1,0){3}{\sfr}
\multiput(.3,1.3)(1,0){3}{\sfr}
\put(.3,.3){\sfr}
\end{picture}}}

\newcommand{\sthreethreetwobox}
{\stln \lower3.8ex\hbox{
\begin{picture}(3.6,3.6)
\multiput(.3,2.3)(1,0){3}{\sfr}
\multiput(.3,1.3)(1,0){3}{\sfr}
\put(.3,.3){\sfr}
\put(1.3,.3){\sfr}
\end{picture}}}

\newcommand{\sthreethreethreebox}
{\stln \lower3.8ex\hbox{
\begin{picture}(3.6,3.6)
\multiput(.3,2.3)(1,0){3}{\sfr}
\multiput(.3,1.3)(1,0){3}{\sfr}
\multiput(.3,.3)(1,0){3}{\sfr}
\end{picture}}}

\newcommand{\sgencolbox}
{\stln \lower8.6ex\hbox{
\begin{picture}(1.6,7.6)
\multiput(.3,4.3)(0,1){3}{\sfr}
\put(.3,1.3){\framebox(1,3){$\vdots$}}
\put(.3,.3){\sfr}
\end{picture}}}

\newcommand{\sgenrowonebox}
{\stln \lower2.6ex\hbox{
\begin{picture}(7.6,2.6)
\multiput(.3,1.3)(1,0){3}{\sfr}
\put(3.3,1.3){\framebox(3,1){$\cdots$}}
\put(6.3,1.3){\sfr}
\put(.3,.3){\sfr}
\end{picture}}}

\newcommand{\sgenrowtwobox}
{\stln \lower2.6ex\hbox{
\begin{picture}(7.6,2.6)
\multiput(.3,1.3)(1,0){3}{\sfr}
\put(3.3,1.3){\framebox(3,1){$\cdots$}}

\put(6.3,1.3){\sfr}
\put(.3,.3){\sfr}
\put(1.3,.3){\sfr}
\end{picture}}}

\newcommand{\marcshapirowbox}
{\stln \lower2.6ex\hbox{
\begin{picture}(12.6,2.6)
\multiput(.3,1.3)(1,0){3}{\sfr}
\put(3.3,1.3){\framebox(3,1){$\cdots$}}
\put(6.3,1.3){\sfr}
\put(7.3,1.3){\sfr}
\put(8.3,1.3){\framebox(3,1){$\cdots$}}
\put(11.3,1.3){\sfr}
\put(7.3,1){$\underbrace{~~~~~~~~~~~~~~~}_{k}$}
\multiput(.3,.3)(1,0){3}{\sfr}
\put(3.3,.3){\framebox(3,1){$\cdots$}}
\put(6.3,.3){\sfr}
\put(.3,0){$\underbrace{~~~~~~~~~~~~~~~~~~~~~}_{n}$}
\end{picture}}}

\newcommand{\sgentworowonerowbox}
{\stln \lower2.6ex\hbox{
\begin{picture}(11.6,2.6)
\multiput(.3,1.3)(1,0){3}{\sfr}
\put(3.3,1.3){\framebox(3,1){$\cdots$}}
\put(6.3,1.3){\sfr}
\put(7.3,1.3){\framebox(3,1){$\cdots$}}
\put(10.3,1.3){\sfr}
\multiput(.3,.3)(1,0){3}{\sfr}
\put(3.3,.3){\framebox(3,1){$\cdots$}}
\put(6.3,.3){\sfr}
\end{picture}}}

\newcommand{\soneoneoneonebox}
{\stln \lower5ex\hbox{
\begin{picture}(1.6,4.6)
\multiput(.3,.3)(0,1){4}{\sfr}
\end{picture}}}

\newcommand{\soneoneoneoneoneonebox}
{\stln \lower7.4ex\hbox{
\begin{picture}(1.6,6.6)
\multiput(.3,.3)(0,1){6}{\sfr}
\end{picture}}}

\newcommand{\stwotwooneonebox}
{\stln \lower5ex\hbox{
\begin{picture}(2.6,4.6)
\multiput(.3,.3)(0,1){4}{\sfr}
\put(1.3,2.3){\sfr}
\put(1.3,3.3){\sfr}
\end{picture}}}

\newcommand{\sgenrowrowbox}
{\stln \lower2.6ex\hbox{
\begin{picture}(7.6,2.6)
\multiput(.3,1.3)(1,0){3}{\sfr}
\put(3.3,1.3){\framebox(3,1){$\cdots$}}
\put(6.3,1.3){\sfr}
\multiput(.3,.3)(1,0){3}{\sfr}
\put(3.3,.3){\framebox(3,1){$\cdots$}}
\put(6.3,.3){\sfr}
\end{picture}}}

\newcommand{\eq}{\begin{equation}}
\newcommand{\en}{\end{equation}}
\newcommand{\eqn}{\begin{eqnarray}}
\newcommand{\enn}{\end{eqnarray}}
\newcommand{\nn}{\nonumber }
\newcommand{\beq}{\begin{equation}}
\newcommand{\eeq}{\end{equation}}
\begin{document}
\begin{titlepage}
\begin{flushright}
  PSU-TH-205\\
\end{flushright}
\begin{center}
{\bf NOVEL SUPERMULTIPLETS OF $SU(2,2,|4)$ AND THE $AdS_{5}/CFT_{4}$
DUALITY } \\
\vspace{1cm}
{\bf M. G\"{u}naydin\footnote{Work supported in part by the
National Science Foundation under Grant Number PHY-9802510. \newline
e-mail: murat@phys.psu.edu}
, D. Minic\footnote{
e-mail: minic@physics.usc.edu. Present address : Department of
Physics and Astronomy, University of Southern California, Los Angeles,
CA 90089-0484}
and M. Zagermann} \footnote{
e-mail: zagerman@phys.psu.edu}  \\
\vspace{.5cm}
Physics Department \\
Penn State University\\
University Park, PA 16802 \\
\vspace{.5cm}
{\bf Abstract}
\end{center}
We continue our study of  the unitary supermultiplets of the ${\cal{N}}=8$ 
$d=5$ 
anti-de Sitter ($AdS_5$)
superalgebra $SU(2,2|4)$, which is the symmetry group of type IIB superstring 
theory
on $AdS_5 \times S^5$. $SU(2,2|4)$ is also the ${\cal{N}}=4$ extended conformal
superalgebra in $d=4$. We show explicitly how to go from the compact
$SU(2)\times SU(2)\times U(1)$ basis to the non-compact $SL(2,\mathbf{C})
\times \mathcal{D}$ basis of the positive (conformal) energy unitary  
representations of the conformal group $SU(2,2)$ in $d=4$. The doubleton
representations of the $AdS_5$ group $SU(2,2)$, which do not have a smooth 
Poincar\'{e} limit in $d=5$, are shown to represent fields with vanishing
masses in four dimensional Minkowski space, i.e. on the boundary of
$AdS_5$, where $SU(2,2)$ acts as conformal group. The unique CPT 
self-conjugate irreducible doubleton supermultiplet of $SU(2,2|4)$
is
simply the ${\cal{N}}=4$ Yang-Mills supermultiplet in $d=4$. We study 
some novel  short non-doubleton supermultiplets of $SU(2,2|4)$ that have spin range 2 and that do not appear in  the Kaluza-Klein spectrum of type 
IIB supergravity or in
tensor products of the ${\cal{N}}=4$ Yang-Mills supermultiplet
with itself. These novel supermultiplets can be obtained from tensoring
 chiral doubleton supermultiplets ,some of which we expect to be related to the  
massless limits of 1/4 BPS states. Hence, these novel supermultiplets
may be relevant to the solitonic sector of IIB superstring and/or $(p,q)$ superstrings over $AdS_5 \times S^5$.

\end{titlepage}

\renewcommand{\theequation}{\arabic{section} - \arabic{equation}}
\section{Introduction}
\setcounter{equation}{0}
This past year a great deal of work has been done on AdS/CFT
(anti-de Sitter/conformal field theory) dualities in various
dimensions. This recent interest in AdS/CFT dualities 
was primarily started with
the original conjecture of Maldacena \cite{mald}
relating the large $N$ limits of certain conformal field theories
in $d$ dimensions to M-theory/string theory
compactified to $d+1$-dimensional AdS spacetimes. Maldacena's
conjecture was motivated by the work on 
properties of the physics of $N$ $Dp$-branes in the
near horizon limit \cite{ads} and the much earlier work on
10-$d$ IIB supergravity
compactified on $AdS_5 \times S^5$ and
11-$d$ supergravity compactified on $AdS_7 \times S^4$ and
$AdS_4 \times S^7$\cite{dfhn, mgnm, krv, ptn, mgnw, gnw, ss}.
Maldacena's conjecture was made
more precise in \cite{pol,witt}.

The relation between Maldacena's
conjecture and the dynamics of the singleton and doubleton fields that
live on the boundary of AdS 
spacetimes was reviewed in \cite{sfcf,mgdm} and its relation
to  the spectra of maximal supergravities in eleven and 
ten dimensions in \cite{mgdm,dupo}.

The prime example of this
AdS/CFT duality is the duality between the large $N$ limit
of ${\cal{N}}=4$ $SU(N)$ super Yang-Mills theory in $d=4$ and type IIB 
superstring theory on $AdS_5 \times S^5$. In our earlier work \cite{gmz1} 
we studied
the unitary supermultiplets of the ${\cal{N}}=8$ $d=5$ 
anti-de Sitter 
superalgebra $SU(2,2|4)$ and  gave a complete classification of the doubleton
supermultiplets of $SU(2,2|4)$. The doubleton supermultiplets do not have 
a smooth Poincar\'{e} limit in $d=5$. They correspond
to $d=4$ superconformal field theories  living on the boundary of 
$AdS_5$, 
where $SU(2,2|4)$ acts as the ${\cal{N}}=4$ extended superconformal algebra.
The unique CPT self-conjugate irreducible 
doubleton supermultiplet is simply 
the  ${\cal{N}}=4$ super Yang-Mills multiplet in $d=4$ \cite{mgnm}. However, there
are also chiral (i.e. non-CPT self-conjugate) doubleton supermultiplets 
with higher spins. The maximum spin range
of the general doubleton supermultiplets is 2. 
We also studied the 
supermultiplets of $SU(2,2|4)$ that
can be obtained by 
tensoring two doubleton supermultiplets.  This class of supermultiplets 
has a maximal spin range of four and contains the multiplets that
are commonly referred to as ``massless'' in the $5d$ AdS sense including
the ``massless'' ${\cal{N}}=8$ graviton supermultiplet
in $AdS_5$ with   spin range two. Some of these supermultiplets 
were studied recently \cite{ferrara} using the language of
${\cal{N}}=4$  conformal superfields developed sometime ago
\cite{hst}.

In this paper we continue our study of the unitary supermultiplets
of $SU(2,2|4)$ and their relevance to the AdS/CFT duality.
To make contact with the standard language used in conformal field theory,
we first show explicitly in section two below how to go from the compact
$SU(2)\times SU(2)\times U(1)$ basis to the conventional non-compact 
$SL(2,\mathbf{C})
\times \mathcal{D}$ ( Lorentz group times dilatations) basis of the positive (conformal) energy unitary 
representations of the conformal group $SU(2,2)$ in $d=4$.
The compact $SU(2)\times SU(2) \times U(1)$ labels $(j_L,j_R,E)$
are shown to coincide with the non-compact labels ($j_M, j_N, -l$)
of $SL(2, \mathbf{C}) \times \mathcal{D}$. The doubleton
representations of the $AdS_5$ group $SU(2,2)$, which do not have a smooth 
Poincar\'{e} limit in five dimensions, are shown to represent fields with 
vanishing
masses in four dimensional Minkowski spacetime, i.e. on the boundary of
$AdS_5$. 
In section 3  we write down the ${\cal{N}}=4$ extended conformal superalgebra
in $d=4$ in a non-compact as well as in a compact basis. 
In section 4 we recapitulate the classification of the doubleton
supermultiplets. In section 5 we study some novel short supermultiplets of  
$SU(2,2|4)$ that have spin range 2. We conclude with a discussion
of the relevance of our results to the AdS/CFT correspondence. In particular,
we point out 
that even though these novel short supermultiplets do not occur in 
tensor products of the ${\cal{N}}=4$ Yang-Mills supermultiplet
with itself, they can be obtained by tensoring of higher spin chiral doubleton
supermultiplets. We argue that massless limits of $1/4$ BPS states in ${\cal{N}}=4$
super Yang-Mills theory \cite{ob} must involve chiral spin $3/2$ doubleton supermultiplets.  This implies that the novel short supermultiplets, that 
 do not appear in the Kaluza-Klein spectrum of type
IIB supergravity, may be relevant to the solitonic sector of IIB superstring
and/or $(p,q)$ superstrings over $AdS_5 \times S^5$.

\section{Compact ($SU(2)\times SU(2)\times U(1))$ versus non-compact
$(SL(2,\mathbf{C}) \times \mathcal{D})$ bases for the positive energy unitary
representations of the  group $SU(2,2)$ }
\setcounter{equation}{0}

The conformal group in four dimensions $SU(2,2)$ (the two sheeted 
covering of $SO(4,2)$) is generated by the Lorentz group 
generators $M_{\mu\nu}$, the four momentum $P_{\mu}$, the dilatation 
generator $D$ and the generators 
of  special conformal transformations $K_{\mu}$ ($\mu, \nu, 
\dots = 0, 1, 2, 3$). The commutation relations are
\begin{eqnarray}
[M_{\mu\nu},M_{\rho\sigma}] & = &i (\eta_{\nu\rho}M_{\mu\sigma}-
\eta_{\mu\rho}M_{\nu\sigma} -\eta_{\nu\sigma}M_{\mu\rho}+
\eta_{\mu\sigma}M_{\nu\rho})\cr
[ P_{\mu}, M_{\rho\sigma} ] & = & i (\eta_{\mu\rho}P_{\sigma}-\eta_{\mu\sigma}
P_{\rho})\cr
[K_{\mu},M_{\rho\sigma}]& = &i (\eta_{\mu\rho}K_{\sigma}-\eta_{\mu\sigma}
K_{\rho})\cr
[D,M_{\mu\nu}]& = & [P_{\mu},P_{\nu}] = [K_{\mu},K_{\nu}]=0\cr
[P_{\mu},D] & = &iP_{\mu}; \quad [K_{\mu},D]=-iK_{\mu}\cr
[P_{\mu},K_{\nu}]& = &2i(\eta_{\mu\nu}D-M_{\mu\nu})
\end{eqnarray}
with $\eta_{\mu\nu}=\textrm{diag}(+,-,-,-)$.

Defining
\eq
M_{\mu 5} = {1 \over 2} (P_{\mu} - K_{\mu}), \quad
M_{\mu 6} = {1 \over 2} (P_{\mu} + K_{\mu}), \quad M_{56} = -D,
\en
the isomorphy to $SO(4,2)$ becomes manifest
($-\eta_{55}=\eta_{66}=1$; $\quad a,b,\dots=0,1,2,3,5,6$)
\eq
[M_{ab}, M_{cd}] = i(\eta_{bc}M_{ad} - \eta_{ac}M_{bd}
-\eta_{bd}M_{ac} + \eta_{ad}M_{bc}).\label{SO42}
\en
Considering $SU(2,2)$ as the isometry group of five dimensional 
anti-de Sitter space, the above generators
have a different physical interpretation. In particular, the rotation
group becomes $Spin(4)= SU(2) \times SU(2)$, generated by $M_{mn}$ and 
$M_{m5}$ ($m, n, \dots = 1,2,3$).  The  generator $E\equiv M_{06}$
becomes the AdS energy generating translations along the timelike Killing
vector field of $AdS_{5}$, and the non-compact
generators $M_{0m}$, $M_{m6}$ correspond to ``boosts'' and spacelike 
``translations'' in $AdS_{5}$.
  
There are two different subgroups of the conformal group which play an 
important role in the classification of its physically relevant 
representations:

i) The maximal compact subgroup $SU(2)_{L}\times SU(2)_{R}\times U(1)_{E}$
generated by the compact generators $M_{mn}$, $M_{m5}$, $E\equiv M_{06}$, 
which after being relabelled as
\eqn
L_{m}&=&\frac{1}{2} \left( \frac{1}{2} \varepsilon_{mnl} M_{nl}+M_{m5}\right)
\qquad 
\longrightarrow SU(2)_{L}\nn\cr
R_{m}&=&\frac{1}{2} \left( \frac{1}{2} \varepsilon_{mnl} M_{nl}-M_{m5}\right)
\qquad 
\longrightarrow SU(2)_{R}
\enn
satisfy
\eqn
[L_{m},L_{n}]&=&i \varepsilon_{mnl}L_{l}\nn\cr
[R_{m},R_{n}]&=&i \varepsilon_{mnl}R_{l}\nn\cr
[L_{m},R_{n}]&=& [E,L_{n}]=[E,R_{n}] =0.
\enn
In the interpretation of $SU(2,2)$ as conformal group, the $U(1)_{E}$ 
generator $E=\frac{1}{2}(P_{0}+K_{0})$ is  simply the 
conformal Hamiltonian. 
Denoting the Lie algebra of $SU(2)_{L}\times SU(2)_{R}\times U(1)_{E}$
by $L^{0}$, the conformal algebra $g$ has a 
three graded decomposition   
\eq
g = L^{+} \oplus L^{0} \oplus L^{-},
\en
where
\eqn
[L^{0},L^{\pm}] &= & L^{\pm} \cr
[L^{+},L^{-}] &=&L^{0} \cr
[L^{+},L^{+}] &=& 0=[L^{-},L^{-}]\cr
[E,L^{\pm}]&=&\pm L^{\pm}, \quad [E,L^{0}]=0.
\enn

ii) The stability group $H$ of $x^{\mu}=0$ when $SO(4,2)$ acts in the usual 
way on 
the (conformal compactification of) $4d$ Minkowski spacetime. Its Lie algebra
consists of the 
generators $ M_{\mu\nu}$ of the Lorentz group $SL(2,\mathbf{C})$, the 
dilatation operator $D$ and the generators of the special conformal 
transformations $K_{\mu}$. Thus $H$ is the semi-direct
product $(SL(2,{\mathbf{C}})\times {\mathcal{D}})\odot {\mathcal{K}}_4 $,
where ${\mathcal{K}}_4$ represents the Abelian subgroup generated by the 
special
conformal generators $K_{\mu}$.
Introducing
\eqn
M_{m}&=&\frac{1}{2} \left( \frac{1}{2} \varepsilon_{mnl} M_{nl}+ i M_{0m}
\right)\nn\cr
N_{m}&=&\frac{1}{2} \left( \frac{1}{2} \varepsilon_{mnl} M_{nl}- i M_{0m}
\right),
\enn
the $SL(2,\mathbf{C})$ part reads
\eqn
[M_{m},M_{n}]&=&i \varepsilon_{mnl}M_{l}\nn\cr
[N_{m},N_{n}]&=&i \varepsilon_{mnl}N_{l}\nn\cr
[M_{m},N_{n}]&=&0.
\enn

Physically relevant representations of the conformal group are unitary 
irreducible representations (UIR's) of the lowest weight type in which  
the spectrum of the conformal Hamiltonian  
(resp. the AdS energy) $E$ is bounded from below.

The most natural way to construct them is to work in a  
$SU(2)_{L}\times SU(2)_{R}\times U(1)_{E}$ covariant basis so that
the lowest weight property can reveal itself in a manifest way. 
The lowest weight UIR's of 
$SU(2,2)$
can then be constructed in a simple way by using the oscillator method of
\cite{mgcs,ibmg,mgnm,gmz1}
and are uniquely determined  by the 
quantum numbers $(j_{L},j_{R},E)$ of an irreducible  
$SU(2)_{L}\times SU(2)_{R}\times U(1)_{E}$ representation 
$|\Omega\rangle$ that is annihilated by the elements of $L^{-}$.

In conformal field theory, on the other hand, the conformal group 
is usually represented   
on fields that live on $4d$ Minkowski spacetime 
and transform covariantly under the Lorentz group $SL(2,\mathbf{C})$ 
and the dilatations. The standard way to construct these fields is via 
the method of induced representations \cite{macksalam}, 
in which a (usually 
finite dimensional) representation of the stability group $H$ induces a 
representation of the whole conformal group $G$ on fields that 
live on the
coset space $G/H$, which in our case is just the conformal compactification
of $4d$ Minkowski spacetime. Consequently, these representations are labelled
by their $SL(2,\mathbf{C})$ quantum numbers $(j_{M},j_{N})$, their conformal 
dimension $l$ and certain matrices $\kappa_{\mu}$ related to their behaviour 
under special conformal transformations $K_{\mu}$.

To translate between these two viewpoints, we will now present the 
oscillator representations of $SU(2,2)$ in a way which makes the 
transition to the 
$SL(2,\mathbf{C})$- and dilatation covariant field representations more 
obvious.

To this end, let $\gamma_{\mu}$ be the ordinary $4d$ gamma matrices
($\{ \gamma_{\mu},\gamma_{\nu}\}=2\eta_{\mu\nu}$) with
$\gamma_{5}=\gamma_{0}\gamma_{1}\gamma_{2}\gamma_{3}$. Then the
matrices

\eqn
\Sigma(M_{\mu\nu}) &:= &\frac{i}{4}\left[\gamma_{\mu},\gamma_{\nu}\right]\nn\cr
\Sigma(M_{\mu 5}) &:= & \frac{i}{2}\gamma_{\mu}\gamma_{5}\nn\cr
\Sigma(M_{\mu 6}) &:= & \frac{1}{2}\gamma_{\mu}\nn\cr
\Sigma(M_{56}) &:= & \frac{1}{2}\gamma_{5}
\enn
generate a four dimensional (non-unitary) irreducible representation of the 
conformal algebra 
(\ref{SO42}).
In the following, we will choose the ``Dirac representation''

\eqn
\gamma^{0}=\gamma_{0}&=&\left(\begin{array}{cc}
\mathbf{1} & 0\\
0 & -\mathbf{1}\end{array}\right)\nn\cr
\gamma^{m}=-\gamma_{m}&=&\left(\begin{array}{cc}
0 & \sigma^{m}\\
-\sigma^{m} & 0\end{array}\right)\nn\cr
\Rightarrow \gamma_{5} &=& 
i \left(\begin{array}{cc}
0 & \mathbf{1}  \\
\mathbf{1} &0 \end{array}\right),
\enn
where $\sigma^m$ are the usual Pauli matrices. With this choice the 
$\Sigma(M_{ab})$ are given by
\eqn
\Sigma(M_{mn}) & = & \frac{1}{2} \varepsilon_{mnl}\left(\begin{array}{cc}
\sigma^{l}& 0\\
0 &\sigma^{l}\end{array}\right)\cr
\Sigma(M_{0m}) & = & -\frac{i}{2}\left(\begin{array}{cc}
0 &\sigma^{m}\\
\sigma^{m}&0\end{array}\right)\cr
\Sigma(M_{05}) & = & \frac{1}{2}\left(\begin{array}{cc}
0 &-\mathbf{1}\\
\mathbf{1}&0\end{array}\right)\cr
\Sigma(M_{m5}) & = & \frac{1}{2}\left(\begin{array}{cc}
\sigma^{m}&0\\
0&-\sigma^{m}\end{array}\right)\cr
\Sigma(M_{06}) & = & \frac{1}{2}\left(\begin{array}{cc}
\mathbf{1}&0\\
0& -\mathbf{1}\end{array}\right)\cr
\Sigma(M_{m6}) & = & \frac{1}{2}\left(\begin{array}{cc}
0 &-\sigma^{m}\\
\sigma^{m}&0\end{array}\right)\cr
\Sigma(M_{56}) & = & \frac{i}{2}\left(\begin{array}{cc}
0 &\mathbf{1}\\
\mathbf{1}&0\end{array}\right).\label{Sigma}
\enn
Of course, this is nothing but the fundamental representation of $SU(2,2)$,
since
\eq
\gamma^{0}\Sigma(M_{ab})=\Sigma(M_{ab})^{\dagger}\gamma^{0}.\label{dagger}
\en

Consider now $P$ copies (or ``generations'') of oscillator 
operators $a^{i}(K)=
a_{i}(K)^{\dagger}$, $b^{r}(K)=
b_{r}(K)^{\dagger}$
\eq
[a_i(K), a^j(L)] = \delta_{i}^{j} \delta_{KL}, \quad
[b_r(K), b^s(L)] = \delta_{r}^{s} \delta_{KL}
\en
with $i,j=1,2$; $r,s=1,2$; $K,L= 1, \dots, P$,
which we now regroup into a ``spinor'' $\Psi$
\eq
\Psi(K) := \left(\begin{array}{c}
a_{1}(K)\cr
a_{2}(K)\cr
-b^{1}(K)\cr
-b^{2}(K)
\end{array}\right)
\en
so that
\eq
\bar{\Psi}(K)\equiv {\Psi}^{\dagger}(K)\gamma^{0}=\left(a^{1}(K),a^{2}(K),
b_{1}(K), b_{2}(K)\right).
\en

Defining 
\eq
\bar{\Psi}\Sigma(M_{ab})\Psi := 
\sum_{K=1}^{P} \bar{\Psi}(K)\Sigma(M_{ab})\Psi(K), 
\en
one finds
\eq
\left[ \bar{\Psi}\Sigma(M_{ab})\Psi,\bar{\Psi}\Sigma(M_{cd})\Psi\right]=
\bar{\Psi}\left[ \Sigma(M_{ab}),\Sigma(M_{cd})\right]\Psi,
\en
i.e. the $\bar{\Psi}\Sigma(M_{ab})\Psi$ generate an infinite dimensional 
representation of $SU(2,2)$ in the Fock space of the oscillators 
$a^{i}$ and $b^{r}$, which, in
contrast to the finite dimensional representation $\Sigma(M_{ab})$, is now 
unitary because of (\ref{dagger}) and the Hermiticity of $\gamma^{0}$.
A short look at (\ref{Sigma}) reveals that all non-compact generators are
represented by linear combinations of di-creation and di-annihilation 
operators of the form ${\vec{a}}^{i}\cdot {\vec{b}}^{r}$ and 
${\vec{a}}_{i}\cdot {\vec{b}}_{r}$ (Here and in the following, the dot 
product denotes summation over the ``generation'' index $K$, i.e.
${\vec{a}}^{i}\cdot {\vec{b}}^{r}\equiv \sum_{K=1}^{P} a^{i}(K)b^{r}(K)$, 
etc.). 
As for the compact generators,  
one finds that
the generators $L_{m}$ of $SU(2)_{L}$ (to simplify the notation we will 
from now on just 
write $M_{ab}$, $L_{m}$,
etc. instead of $\bar{\Psi}\Sigma(M_{ab})\Psi$, 
$\bar{\Psi}\Sigma(L_{m})\Psi$ \dots) are given by linear combinations of
\eq
L^{k}_{i} := {\vec{a}}^{k} \cdot {\vec{a}}_{i}
-{1 \over 2} \delta^{k}_{i}N_{a}, 
\en
whereas the generators $R_{m}$ of $SU(2)_{R}$ are linear combinations of
\eq
R^{r}_{s} := {\vec{b}}^{r} \cdot {\vec{b}}_{s}
-{1 \over 2} \delta^{r}_{s} N_{b}
\en
and $E$ is simply given by
\eq
E = {1 \over 2} (N_a + N_b + 2P),
\en
where $N_{a} \equiv {\vec{a}}^{i} \cdot {\vec{a}}_{i}$, 
$N_{b} \equiv {\vec{b}}^{r} \cdot {\vec{b}}_{r}$ are the
bosonic number operators,
in complete agreement with the construction in \cite{gmz1}.

As mentioned above, the positive energy UIR's are then obtained by 
constructing an irreducible representation $|\Omega\rangle$ of 
$SU(2)_{L}\times SU(2)_{R}\times U(1)_{E}$ in the Fock space of the 
oscillators  with quantum 
numbers $(j_{L},j_{R},E)$ that is annihilated by all the generators 
${\vec{a}}_{i}\cdot {\vec{b}}_{r}$ of $L^{-}$:
\eq
\vec{a}_{i}\cdot\vec{b}_{r}|\Omega\rangle=0.
\en
Acting  repeatedly with the di-creation operators 
$\vec{a}^{i}\cdot\vec{b}^{r}$
of $L^{+}$
on $|\Omega\rangle$, one generates the basis of a positive energy UIR of the 
whole group $SU(2,2)$.

To see the relation to the $SL(2,\mathbf{C})$- and $D$-covariant induced 
representations on fields, consider the operator
\eq
U:=e^{\vec{a}^{1}\cdot\vec{b}^{1}+\vec{a}^{2}\cdot\vec{b}^{2}}.
\en

It has the following important property 
\begin{eqnarray}
M_{m}U & = & U(L_{m}+L^{-})\nonumber\\
N_{m}U & = & U(R_{m}+L^{-})\nonumber\\
D U    & = & U(-iE+L^{-})\nonumber\\
K_{\mu}U & = & U(L^{-}),\label{UL}
\end{eqnarray}
where $L^{-}$ stands for certain linear combinations of di-annihilation 
operators $\vec{a}_{i}\cdot\vec{b}_{r}$.

Acting with $U$ on a lowest weight vector $|\Omega \rangle$ corresponds 
to a rotation in the corresponding representation space of $SU(2,2)$:
\eq
U|\Omega \rangle = e^{\bar{\Psi} \Sigma (M_{05}+iM_{56}) \Psi} |\Omega \rangle.
\en
Since $L^{-}|\Omega\rangle=0$, it then follows from (\ref{UL}) that $\Phi(0):=
U|\Omega\rangle$ is an
irreducible representation of the little group $H$ with $SL(2,\mathbf{C})$
quantum numbers $(j_{M},j_{N})=(j_{L},j_{R})$, conformal dimension\footnote{
In our conventions, $l$ is the length (or inverse mass) dimension.} 
$l=-E$
and trivially represented special conformal transformations $K_{\mu}$ 
(i.e. $\kappa_{\mu}=0$). 
Acting with $e^{-i x^{\mu}P_{\mu}}$ on $\Phi(0)$ then translates the 
field in Minkowski space:
\eq
\Phi(x^{\mu})=e^{-i x^{\mu}P_{\mu}}\cdot \Phi(0) =
e^{-i x^{\mu}P_{\mu}}U|\Omega\rangle
\en
and generates a (group theoretically equivalent) induced representation 
of $SU(2,2)$ along the 
lines of \cite{macksalam} with exact numerical coincidence of the compact 
and the covariant labels $(j_{L},j_{R},E)$ and $(j_{M},j_{N},-l)$.
Since the bosonic oscillators in terms of which we realized the generators
transform in the spinor representation of $SU(2,2)$, the oscillator
realization could be reinterpreted in the language of twistors.

To conclude this section, we finally note that the lowest weight UIR's 
of $SU(2,2)$ with vanishing $4d$ mass\footnote{Note that we are talking here 
about 
the mass in the \emph{four} dimensional sense and not about mass in five 
dimensional $AdS$ space, where already an invariant definition of this quantity
poses a problem \cite{gmz1}.} 
$m^{2}=P_{\mu}P^{\mu}$ 
are exactly
the ones obtained by taking only one generation of oscillators (i.e. for $P=1$)
\cite{binegar},
since 
\eq
P_{\mu}P^{\mu}=\left[({\vec{c}}^{ \;1}\cdot {\vec{d}}^{ \;1})({\vec{c}}^{ \;2}
\cdot{\vec{d}}^{ \;2})-
({\vec{d}}^{ \;1}\cdot {\vec{c}}^{ \;2})({\vec{d}}^{ \;2}
\cdot {\vec{c}}^{ \;1})\right]
\en
with the (mutually commuting) operators
\eqn
c^{1}(K)&:=&(b_{1}(K)+a^{1}(K))\nn\cr
c^{2}(K)&:=&(b_{2}(K)+a^{2}(K))\nn\cr
d^{1}(K)&:=&(a_{1}(K)+b^{1}(K))\nn\cr
d^{2}(K)&:=&(a_{2}(K)+b^{2}(K))
\enn
vanishes identically in this case.

If $SU(2,2)$ is interpreted as the isometry group of $AdS_{5}$, these 
representations are just the doubleton representations, which do not have a 
smooth $5d$ Poincar\'{e} limit and live on the 
$4d$ boundary of anti-de Sitter space (where they are thus massless 
representations of the $4d$ conformal group solving the free wave equation).
In terms of the oscillator method, they correspond to lowest weight
vectors of the form $|\Omega\rangle=a^{i_{1}}\ldots a^{i_{n}}|0\rangle$
or $|\Omega\rangle=b^{r_{1}}\ldots b^{r_{n}}|0\rangle$ with $n\geq 0$ leading 
to $(j_{L},j_{R},E)=(n/2,0,n/2+1)$ or $(0, n/2, n/2+1)$, respectively.

\section{The superalgebra $SU(2,2|4)$ }
\setcounter{equation}{0}

The centrally extended symmetry supergroup of type IIB superstring theory on $AdS_{5}\times S^{5}$
is the supergroup $SU(2,2|4)$ with the even subgroup
$SU(2,2) \times SU(4) \times U(1)_Z$, where $SU(4)$ is the double cover of 
$SO(6)$, the isometry group of the five sphere \cite{mgnm}. The Abelian $U(1)_Z$
generator, which we will call $Z$,
commutes with all the other
generators and acts like a central charge. Therefore, $SU(2,2|4)$
is not a simple Lie superalgebra. By factoring out this Abelian
ideal one obtains a simple Lie superalgebra, denoted by $PSU(2,2|4)$,
whose even subalgebra is simply $SU(2,2)\times SU(4)$.
For its possible applications to the full  superstring/M-theory 
we consider below the centrally extended supergroup $SU(2,2|4)$
as we did in our earlier paper \cite{gmz1}. The representations 
of $PSU(2,2|4)$ correspond simply to representations of $SU(2,2|4)$
with $Z=0$. We should also note that
both $SU(2,2|4)$ and $PSU(2,2|4)$ admit an outer automorphism $U(1)_Y$
that can be identified with a $U(1)$ subgroup of the $SU(1,1)_{global} \times U(1)_{local}$ symmetry of IIB supergravity in $d=10$ \cite{mgnm}.
$SU(2,2|4)$ can be interpreted as the ${\cal{N}}=8$ extended AdS superalgebra
in $d=5$ or as the ${\cal{N}}=4$ extended conformal superalgebra in $d=4$.

The algebra of ${\cal{N}}$-extended conformal supersymmetry  in $d=4$  can 
be written in a covariant  form as follows 
($i,j =1, \ldots ,{\cal{N}}$; $a,b =0,1,2,3,5,6$)\cite{sohn}:
\begin{eqnarray}
[\Xi_{i}, M_{ab}] =  \Sigma(M_{ab}) \Xi_{i},& &
[{\bar{\Xi}}^{i}, M_{ab}] = - {\bar{\Xi}}^{i}\Sigma(M_{ab})\nonumber\\
\{ \Xi_{i}, \Xi_{j} \} = 
\{ {\bar{\Xi}}^{i}, {\bar{\Xi}}^{j} \} =0, & &
\{ \Xi_{i}, {\bar{\Xi}}^{j} \} = 
2{\delta}^{j}_{i} \Sigma(M^{ab}) M_{ab} -4B^j_i\nonumber\\
{[}B^j_i, M_{ab}{]}=0, & & {[}B^j_i, B^l_k{]}={\delta}^l_i B^j_k - 
{\delta}^j_k B^l_i\nonumber\\
{[}\Xi_{i}, B^k_j{]} = {\delta}^k_i \Xi_{j}  - {1 \over 4} 
{\delta}^k_j \Xi_{i}, & & 
{[}{\bar{\Xi}}^{i}, B^k_j{]} = -{\delta}^i_j {\bar{\Xi}}^{k}  + {1 \over 4} 
{\delta}^k_j {\bar{\Xi}}^{i},
\end{eqnarray}
where the (four component) conformal spinor $\Xi$ is defined in terms of the
the chiral components of the Lorentz spinors $Q$ and $S$ (the generators of
Poincar\'{e} and $S$ type supersymmetry) as
\eq
\Xi \equiv \left(\matrix{Q_{\alpha} \cr
                        {\bar{S}}^{\dot{\alpha}} \cr} \right).
\en
The $B_i^j$ are the generators of the internal (R-)symmetry group 
$U({\cal{N}})$ and the  $\Sigma(M_{ab})$ are the $4 \times 4$ matrices 
introduced in the previous section.

The superalgebra $SU(2,2|4)$ has a three graded decomposition with
respect to its compact subsuperalgebra $SU(2|2)\times SU(2|2) \times U(1)$
\eq
g = L^{+} \oplus L^{0} \oplus L^{-},
\en
where
\eqn
[L^{0},L^{\pm}] & \subseteq & L^{\pm} \cr
[L^{+},L^{-}] & \subseteq &L^{0} \cr
[L^{+},L^{+}] &=& 0=[L^{-},L^{-}].
\enn
Here $L^{0}$ represents the generators of
$SU(2|2) \times SU(2|2) \times U(1)$.

Generalizing the (purely bosonic) oscillator construction for $SU(2,2)$ in
section 2,
the Lie superalgebra $SU(2,2|4)$
can be realized in terms of bilinear combinations of bosonic and
fermionic annihilation and creation operators $\xi_{A}$
($\xi^{A}={\xi_{A}}^{\dagger}$) and $\eta_{M}$
($\eta^{M}={\eta_{M}}^{\dagger}$)
which transform covariantly and contravariantly
under the  two $SU(2|2)$ subsupergroups of $SU(2,2|4)$ 
\cite{ibmg,mg81,mgnm,gmz1}
\eq
\xi_{A} = \left(\matrix{a_{i} \cr
                        \alpha_{\gamma} \cr} \right) ,\quad
\xi^{A} = \left(\matrix{a^{i} \cr
                        \alpha^{\gamma} \cr} \right)
\en
and
\eq
\eta_{M} = \left(\matrix{b_{r} \cr
                        \beta_{x} \cr} \right) , \quad
\eta^{M} = \left(\matrix{b^{r} \cr
                        \beta^{x} \cr} \right)
\en
with $i,j=1,2$; $\gamma,\delta=1,2$; $r,s=1,2$; $x,y=1,2$ and
\eq
[a_i, a^j] = \delta_{i}^{j} , \quad
\{\alpha_{\gamma}, \alpha^{\delta}\} = \delta_{\gamma}^{\delta}
\en
\eq
[b_r, b^s] = \delta_{r}^{s} , \quad
\{\beta_{x}, \beta^{y}\} = \delta_{x}^{y}.
\en
Again,  annihilation and creation operators are labelled by lower and
upper indices, respectively. 
The generators of $SU(2,2|4)$ are given in terms of the above 
superoscillators schematically as
\eqn
L^{-} &=& {\vec{\xi}}_{A} \cdot {\vec{\eta}}_{M} \cr
L^{0} &=& {\vec{\xi}}^{A} \cdot {\vec{\xi}}_{B}
\oplus {\vec{\eta}}^{M} \cdot {\vec{\eta}}_{N} \cr
L^{+} &=& {\vec{\xi}}^{A} \cdot {\vec{\eta}}^{M},\label{xieta}
\enn
where the arrows over $\xi$ and $\eta$ again indicate that we are taking an
arbitrary number $P$ of ``generations'' of
superoscillators and the dot represents the summation
over the internal index $K = 1,...,P$, i.e
${\vec{\xi}}_{A} \cdot {\vec{\eta}}_{M} \equiv \sum_{K=1}^{P}
{\xi_{A}}(K){\eta_{M}}(K)$.

The even subgroup $SU(2,2)\times SU(4)\times U(1)_Z$ is obviously generated 
by the di-bosonic and di-fermionic generators. In particular, one
recovers the  $SU(2,2)$ generators of section 2 in terms of the bosonic 
oscillators:
\eqn
L^{j}_{i} &=& {\vec{a}}^{j} \cdot {\vec{a}}_{i}
-{1 \over 2} \delta^{j}_{i}N_{a} \cr
R^{r}_{s} &=& \vec{b^{r}} \cdot \vec{b_{s}}
-{1 \over 2} \delta^{r}_{s} N_{b}\cr
E &=& \frac{1}{2} \{ {\vec{a}}^{i} \cdot {\vec{a}}_{i} + {\vec{b}}_{r} 
\cdot {\vec{b}}^{r}\}=
{1 \over 2} \{ N_a +  N_b +2P  \}\cr
L^{-}_{ir} &=&{\vec{a}}_{i} \cdot {\vec{b}}_{r}, \quad
L^{+ ri}={\vec{b}}^{r} \cdot {\vec{a}}^{i}\label{su22}
\enn
satisfying
\eq
[L^{-}_{ir},L^{+ sj}]
= \delta^{s}_{r} L^{j}_{i} + \delta^{j}_{i} R^{s}_{r}
+ \delta^{j}_{i}\delta^{s}_{r} E. 
\en
Here, $N_{a} \equiv {\vec{a}}^{i} \cdot {\vec{a}}_{i}, 
N_{b} \equiv {\vec{b}}^{r} \cdot {\vec{b}}_{r}$ are again the
bosonic number operators.

Similarly, the $SU(4)$ generators in their $SU(2)\times SU(2) \times U(1)$ 
basis are expressed  in 
terms of the fermionic oscillators $\alpha$ and $\beta$:
\eqn
A^{\delta}_{\gamma} &=& {\vec{\alpha}}^{\delta} \cdot {\vec{\alpha}}_{\gamma}
-{1 \over 2} \delta^{\delta}_{\gamma} N_{\alpha} \cr
B^{y}_{x} &=& {\vec{\beta}}^{y} \cdot {\vec{\beta}}_{x}
-{1 \over 2} \delta^{y}_{x} N_{\beta} \cr
C &=& \frac{1}{2}\{-{\vec{\alpha}}^{\delta} \cdot {\vec{\alpha}}_{\delta}
+ \vec{\beta_{x}} \cdot {\vec{\beta}}^{x}\}={1 \over 2} \{- N_{\alpha} -  
N_{\beta} +2P\} \cr
L^{-}_{\gamma x} & = &{\vec{\alpha}}_{\gamma} \cdot {\vec{\beta}}_{x},
\quad L^{+ x\gamma}={\vec{\beta}}^{x} \cdot 
{\vec{\alpha}}^{\gamma}\label{su4}
\enn
with the closure relation
\eq
[L^{-}_{\gamma x},L^{+ y \delta}]
= -\delta^{y}_{x} A^{\delta}_{\gamma} - \delta^{\delta}_{\gamma} B^{y}_{x}
+ \delta^{y}_{x} \delta^{\delta}_{\gamma} C.
\en
Here $N_{\alpha}={\vec{\alpha}}^{\delta} \cdot {\vec{\alpha}}_{\delta} $ 
and $N_{\beta} = \vec{\beta^{x}} \cdot {\vec{\beta}}_{x}$
are the fermionic number operators.

Finally, the central charge-like $U(1)_Z$ generator $Z$ is given by
\eq 
Z=\frac{1}{2}\{ N_{a}+N_{\alpha}-N_{b}-N_{\beta} \}.
\en
Analogously, the odd generators are given by products of bosonic and 
fermionic oscillators and satisfy the following closure relations  
\eqn
\{ {\vec{a}}_{i} \cdot {\vec{\beta}}_{x},{\vec{\beta}}^{y}
\cdot {\vec{a}}^{j} \}
&=& \delta^{y}_{x} L^{j}_{i} - \delta^{j}_{i} B^{y}_{x}
+ {1 \over 2}\delta^{y}_{x}\delta^{j}_{i} {(E+C+Z)} \cr
\{ {\vec{\alpha}}_{\gamma} \cdot {\vec{b}}_{r},
{\vec{b}}^{s} \cdot {\vec{\alpha}}^{\delta} \}
&=& -\delta^{s}_{r} A^{\delta}_{\gamma} + \delta^{\delta}_{\gamma} R^{s}_{r}
+ {1 \over 2}\delta^{s}_{r} \delta^{\delta}_{\gamma} {(E+C-Z)} \cr
\{ {\vec{a}}^{i} \cdot {\vec{\alpha}}_{\gamma},{\vec{\alpha}}^{\delta}
\cdot {\vec{a}}_{j}\}
&=& \delta^{\delta}_{\gamma} L^{i}_{j} + \delta^{i}_{j} A^{\delta}_{\gamma}
+ {1 \over 2}\delta^{\delta}_{\gamma} \delta^{i}_{j} {(E-C+Z)} \cr
\{ {\vec{b}}^{r} \cdot {\vec{\beta}}_{x},{\vec{\beta}}^{y}
\cdot {\vec{b}}_{s} \}
&=& \delta^{y}_{x} R^{r}_{s} + \delta^{r}_{s} B^{y}_{x}
+ {1 \over 2}\delta^{y}_{x}\delta^{r}_{s} {(E-C-Z)}.
\enn
 
The generator $Y$ of the outer automorphism group $U(1)_Y$ is simply
\eq
Y= N_{\alpha}  - N_{\beta}
\en

\section{Unitary Supermultiplets of $SU(2,2|4)$}
\setcounter{equation}{0}

To construct a basis for a lowest weight UIR of $SU(2,2|4)$,
one starts from a set of states, collectively denoted by $ |\Omega \rangle$, 
in the  Fock space of the oscillators $a$, $b$, $\alpha$, $\beta$ that
transforms irreducibly under $SU(2|2) \times
SU(2|2) \times U(1)$ and that is annihilated by all the generators 
${\vec{\xi}}_{A}\cdot {\vec{\eta}}_{M}\equiv ({\vec{a}}_{i} \cdot {\vec{b}}_{r}
\oplus {\vec{a}}_{i}~\cdot~{\vec{\beta}}_{x}
\oplus {\vec{\alpha}}_{\gamma}~\cdot~{\vec{b}}_{r}
\oplus {\vec{\alpha}}_{\gamma} \cdot {\vec{\beta}}_{x})$ of $L^{-}$
\eq
{\vec{\xi}}_{A}\cdot {\vec{\eta}}_{M}|\Omega\rangle = 0.\label{lwv}
\en 
By acting
on  $ |\Omega \rangle$ repeatedly with $L^{+}$, 
one then
generates an infinite set of states that form a UIR of $SU(2,2|4)$

\eq
|\Omega \rangle ,\quad  L^{+1}|\Omega \rangle ,\quad
L^{+1} L^{+1}|\Omega \rangle , ...
\en
The irreducibility of the resulting representation of $SU(2,2|4)$ follows 
from the irreducibility of $|\Omega \rangle$ under $SU(2|2) \times
SU(2|2) \times U(1)$.
Because of the property (\ref{lwv}), $|\Omega\rangle$ 
as a whole
will be referred to as the ``lowest weight vector (lwv)'' of the
corresponding UIR of $SU(2,2|4)$.

In the restriction to the subspace involving purely bosonic
oscillators, the above construction reduces to the subalgebra $SU(2,2)$ 
and its positive energy UIR's as described in section 2. 
Similarly, when restricted to the subspace involving purely
fermionic oscillators, one gets the compact internal 
symmetry group $SU(4)$ (\ref{su4}),
and the above construction yields the representations of $SU(4)$ in
its $SU(2)\times SU(2) \times U(1)$ basis.

Accordingly, a lowest weight UIR of $SU(2,2|4)$  decomposes into a 
direct sum of
finitely many positive energy UIR's of $SU(2,2)$ 
transforming in certain representations
of the internal symmetry group $SU(4)$. 
Thus each positive energy UIR of $SU(2,2|4)$
corresponds to a supermultiplet of fields living in $AdS_5$ or on its boundary.
The bosonic and fermionic fields in $AdS_5$ or on its boundary will be denoted
as $\Phi_{(j_{L},j_{R})}(E)$ and $\Psi_{(j_{L},j_{R})}(E)$, respectively.
Interpreted as a UIR of the ${\cal{N}}=4$ conformal superalgebra in $d=4$ each 
lowest weight UIR
of $SU(2,2|4)$ corresponds to a supermultiplet of massless or massive
fields. These four dimensional fields  can then be labelled as 
$\Phi_{(j_{M},j_{N})}(l)$ or $\Psi_{(j_{M},j_{N})}(l)$ if they are bosons 
or fermions of conformal dimension $l$ and $SL(2,\mathbf{C})$ quantum numbers
$(j_{M},j_{N})$.

\subsection{Doubleton Supermultiplets of $SU(2,2|4)$}
\setcounter{equation}{0}

By choosing one pair of super oscillators ($\xi$ and $\eta$) in the
oscillator realization of $SU(2,2|4)$ (i.e. for $P=1$), one 
obtains the so-called 
doubleton supermultiplets. The doubleton supermultiplets contain only doubleton
representations of $SU(2,2)$, i.e. they are multiplets of fields living on 
the boundary of $AdS_{5}$ without a $5d$ Poincar\'{e} limit. Equivalently,
they can be characterized as multiplets of massless fields in $4d$ 
Minkowski space that form a UIR of the ${\cal{N}}=4$ superconformal algebra 
$SU(2,2|4)$.

The complete list of doubleton supermultiplets has been given in our previous
paper \cite{gmz1}. In this subsection we shall recapitulate our results
as a preparation for the next section.

The supermultiplet defined by the lwv $|\Omega \rangle = |0\rangle$ 
of $SU(2,2|4)$ 
is the unique irreducible CPT self-conjugate  doubleton supermultiplet. 
It is also the 
supermultiplet of ${\cal{N}}=4$ supersymmetric Yang-Mills theory in $d=4$ 
\cite{mgnm}.
The content of this special doubleton supermultiplet is  given in
Table 1. (We will continue to use this form of presenting our
results in what follows.)

\vspace{.2cm}
\begin{center}
\begin{tabular}{|c|c|c|c|c|c|}
\hline
$SU(2,2) \times SU(4)$ lwv &E=-l &($j_{L},j_{R}$)=($j_M,j_N$) & SU(4) &$Y$ &
Field
\\ \hline
$|0\rangle $          &1 & (0,0)     & 6   & 0 &$\Phi_{0,0}$
\\ \hline
$a^i \beta^x |0\rangle$   &3/2  & (1/2,0) & $\bar{4}$ & -1 &$\Psi_{1/2,0}$
\\ \hline
$b^r \alpha^{\gamma}  |0\rangle$        &3/2& (0,1/2)   & 4  &1 &$\Psi_{0,1/2}$
\\ \hline
$a^i a^j  \beta^x \beta^y |0\rangle$ &2& (1,0)  & $\bar{1}$& -2 &$\Phi_{1,0}$
\\ \hline
$b^r b^s \alpha^{\gamma} \alpha^{\delta} |0\rangle$   &2& (0,1) & 1 & 2 &$
\Phi_{0,1}$
\\ \hline
\end{tabular}
\end{center}
Table 1. The doubleton supermultiplet
corresponding to the lwv $|\Omega \rangle = |0\rangle$.
The first column indicates the lowest weight vectors (lwv) of
$SU(2,2) \times SU(4)$. The second column shows the AdS  energies 
$E=(N_{a}+N_{b})/2+P$ respectively the 
conformal dimensions $l$, and the third coloumn
lists the compact ($j_{L},j_{R}$),
or equivalently,  the non-compact labels ($j_M,j_N$) of the corresponding
fields. Also, $Y= N_{\alpha} - N_{\beta}$, whereas
$\Phi $ and $\Psi$ denote bosonic and fermionic fields, respectively. 
This supermultiplet has $Z=0$. 
\vspace{.7cm}

If we take
\eq
|\Omega \rangle = \xi^A |0\rangle \equiv a^i |0\rangle
\oplus \alpha^{\gamma} |0\rangle
= |{\sonebox}, 1\rangle,
\en
we get a supermultiplet of spin range $3/2$. It is shown in Table 2. 
(See the Appendix for a
quick review of the supertableaux notation \cite{bbars}.)

\vspace{.2cm}
\begin{center}
\begin{tabular}{|c|c|c|c|c|c|}
\hline
$SU(2,2) \times SU(4)$ lwv   &E=-l  & ($j_{L},j_{R}$)=($j_M,j_N$)& SU(4)
&$Y$ &Field
\\ \hline
$a^i|0\rangle $                        &3/2& (1/2,0)  & 6   & 0 &$\Psi_{1/2,0}$
\\ \hline
$a^i a^j \beta^x |0\rangle$ &2  & (1,0)  & $\bar{4}$& -1 &$\Phi_{1,0}$
\\ \hline
$a^i a^j a^k \beta^x \beta^y |0\rangle$&5/2&(3/2,0)&
$\bar{1}$&-2&$\Psi_{3/2,0}$
\\ \hline
$\alpha^{\gamma} |0\rangle$               &1  & (0,0) & 4 &1 &$\Phi_{0,0}$
\\ \hline
$b^r \alpha^{\gamma} \alpha^{\delta} |0\rangle$  &3/2& (0,1/2)  & 1& 2&
$\Psi_{0,1/2}$
\\ \hline
\end{tabular}
\end{center}
Table 2. The doubleton supermultiplet
corresponding to the lwv $|\Omega\rangle=\xi^A |0\rangle
= |{\sonebox},1 \rangle  $. It has $Z= \frac{1}{2}$
\vspace{.7cm}

The CPT conjugate supermultiplet to the one listed in Table 2 is obtained by
taking 
\eq
|\Omega \rangle = \eta^A |0\rangle \equiv b^r |0\rangle
\oplus \beta^{x} |0\rangle
= |1,{\sonebox} \rangle
\en
as the lwv. It is displayed  in Table 3.

\vspace{.2cm}
\begin{center}
\begin{tabular}{|c|c|c|c|c|c|}
\hline
$SU(2,2) \times SU(4)$ lwv &E=-l  & ($j_{L},j_{R}$)=($j_M,j_N$)& SU(4) &$Y$ &
Field
\\ \hline
$b^r|0\rangle $  &3/2& (0,1/2)        & 6  & 0 &$\Psi_{0,1/2}$
\\ \hline
$b^r b^s \alpha^{\gamma} |0\rangle$ &2  & (0,1)& ${4}$& 1 &$\Phi_{0,1}$
\\ \hline
$b^r b^s b^t \alpha^{\gamma} \alpha^{\delta}|0\rangle $ &5/2& (0,3/2)  & 1   & 2
&$\Psi_{0,3/2}$
\\ \hline
$\beta^x |0\rangle$      &1  & (0,0) & $\bar{4}$  &-1 &$\Phi_{0,0}$
\\ \hline
$a^i \beta^x \beta^y |0\rangle$  &3/2& (1/2,0) & $\bar{1}$   & -2
&$\Psi_{1/2,0}$
\\ \hline
\end{tabular}
\end{center}
Table 3. The doubleton supermultiplet
corresponding to the lwv $|\Omega\rangle=\eta^A |0\rangle
= |1,{\sonebox} \rangle  $ . It has $Z=-\frac{1}{2}$.
\vspace{.7cm}

These last two doubleton supermultiplets of spin range 3/2 would occur in the 
${\cal{N}}=4$ super Yang-Mills theory if there is a well-defined  conformal 
(i.e. massless) limit of the 
$1/4$ BPS states described in ref \cite{ob}. 
These  $1/4$ BPS multiplets are massive representations of the four 
dimensional ${\cal{N}}=4$ Poincar\'{e} superalgebra
with two central charges, one of them saturating the BPS bound. As such, 
they are equivalent to  massive representations of the ${\cal{N}}=3$ 
Poincar\'{e} superalgebra without central charges. 
The corresponding multiplet with the lowest spin content (see e.g. \cite{jwjb})
contains 14 scalars, 14 spin 1/2 fermions, six vectors and one spin 3/2 
fermion,
giving altogether $2^{6}$ states. If a massless limit of such a multiplet 
exists, it should decompose into two self-conjugate doubleton multiplets
(Table 1) plus one of the form given in Table 2 plus one of the form given 
in Table 3.

The lowest weight vector of a generic doubleton supermultiplet
is of the form
\eq
|\Omega \rangle = \xi^{A_1} \xi^{A_2}...\xi^{A_{2j}} |0\rangle= 
|\underbrace{\sgenrowbox}_{2j}, 1 \rangle 
\en
or of the ``CPT-conjugate'' form
\eq
|\Omega \rangle = \eta^{A_1} \eta^{A_2}...\eta^{A_{2j}} |0\rangle = 
|1, \underbrace{\sgenrowbox}_{2j} \rangle.
\en

For $j \geq 1$, the corresponding general doubleton supermultiplets
are given in tables 4 and 5. (Note that in these tables we assume 
that $j$ takes on integer values. For $j$ half-integer the roles
of $\Phi$ and $\Psi$ are reversed).

\vspace{.2cm}
\begin{center}
\begin{tabular}{|c|c|c|c|c|}
\hline
E =-l                & ($j_{L},j_{R}$)=($j_M,j_N$)     & SU(4)     & $Y$ 
& Field\\ \hline
j+1               & (j,0)               & 6         & 0 &$\Phi_{j,0}$
\\ \hline
j+{3/2}           & (j+1/2,0)           & $\bar{4}$ & -1 &$\Psi_{j+1/2,0}$
\\ \hline
j+{1/2}           & (j-1/2,0)           & 4         & 1 &$\Psi_{j-1/2,0}$
\\ \hline
j+ 2              & (j+1 ,0)            & $\bar{1}$ & -2 &$\Phi_{j+1,0}$
\\ \hline
j                 & (j-1,0)             & 1         & 2 &$\Phi_{j-1,0}$
\\ \hline
\end{tabular}
\end{center}
Table 4. The doubleton supermultiplet
corresponding to the lwv 
$|\Omega \rangle = \xi^{A_1} \xi^{A_2}...\xi^{A_{2j}} |0\rangle
= |\underbrace{\sgenrowbox}_{2j}, 1 \rangle $. It has $Z=j$.
\vspace{.7cm}

\vspace{.2cm}
\begin{center}
\begin{tabular}{|c|c|c|c|c|}
\hline
E =-l                & ($j_{L},j_{R}$)=($j_M,j_N$)     & SU(4)     & $Y$ 
&Field\\ \hline
j+1               & (0,j)               & 6         & 0 &$\Phi_{0,j}$
\\ \hline
j+{3/2}           & (0,j+1/2)           & 4         & 1 &$\Psi_{0,j+1/2}$
\\ \hline
j+{1/2}           & (0,j-1/2)           & $\bar{4}$ & -1 &$\Psi_{0,j-1/2}$
\\ \hline
j+ 2              & (0,j+1)             & 1         & 2 &$\Phi_{0,j+1}$
\\ \hline
j                 & (0,j-1)             & $\bar{1}$ & -2 &$\Phi_{0,j-1}$
\\ \hline
\end{tabular}
\end{center}
Table 5. The doubleton supermultiplet
corresponding to the lwv 
$|\Omega \rangle = \eta^{A_1} \eta^{A_2}...\eta^{A_{2j}} |0\rangle
= |1, \underbrace{\sgenrowbox}_{2j} \rangle $. It has $Z=-j$.
\vspace{.7cm}

\section{Novel short ``massless'' supermultiplets of $SU(2,2|4)$}
\setcounter{equation}{0}

The doubleton supermultiplets described in the last subsection are 
fundamental in the sense that all other lowest weight UIR's of $SU(2,2|4)$ 
occur in the 
tensor product of two or more doubleton supermultiplets. Instead of trying
to identify these irreducible submultiplets in the (in general reducible, but 
not 
fully reducible) tensor products, one simply increases the number $P$ of
oscillator generations so that the tensoring becomes implicit while the 
irreducibility stays manifest.

The simplest class, corresponding to $P=2$, contains the supermultiplets that 
are commonly referred to as ``massless'' in the $5d$ AdS sense. We will 
therefore use this as a name for all supermultiplets that are obtained by 
taking $P=2$ in the oscillator construction despite some problems with the 
notion of ``mass'' in AdS spacetimes\footnote{Of course, this has nothing to 
do with the completely unambiguous concept of masslessness in  $4d$ Minkowski
spacetime used in Sections 2 and 4.1.} \cite{gmz1}.

Completing our study of these ``massless'' supermultiplets in \cite{gmz1},
we will now give a complete list of the allowed $SU(2,2|4)$ lowest weight 
vectors $|\Omega \rangle$ for $P=2$ and then consider those that give rise 
to the novel short supermultiplets in detail.

The condition $L^{-}|\Omega \rangle = 0$ leaves the following possibilities:

\begin{itemize}
\item $|\Omega \rangle = |0\rangle$. This lwv gives rise to the $\mathcal{N}$ =
8 graviton supermultiplet in $AdS_{5}$ and occurs in the tensor product
of two CPT self-conjugate doubleton (i.e. $\mathcal{N}$ = 4 super Yang Mills) 
supermultiplets.

\item  $|\Omega \rangle = \xi^{A_1}(1) \xi^{A_2}(1) ... \xi^{A_{2j}}(1) 
|0\rangle = |\underbrace{\sgenrowbox}_{2j}, 1 \rangle$. The corresponding 
supermultiplets and also their conjugates resulting from

\item $|\Omega \rangle = \eta^{A_1}(1) \eta^{A_2}(1) ... \eta^{A_{2j}}(1) 
|0\rangle = |1, \underbrace{\sgenrowbox}_{2j} \rangle$ have been listed in
tables 8 to 11 of our previous paper \cite{gmz1}. Increasing $j$ leads
to multiplets with higher and higher spins and AdS energies. For $j>3/2$ 
the spin range is always 4.
None of these multiplets can occur in the 
tensor product of two or more self-conjugate doubleton supermultiplets. They
require the chiral doubleton supermultiplets listed in tables 2 to 5.

\item $|\Omega \rangle = \xi^{A_1}(1) \xi^{A_2}(1) ... \xi^{A_{2j_{L}}}(1)
\eta^{B_1}(2) \eta^{B_2}(2) ... \eta^{B_{2j_{R}}} (2)|0\rangle$\\ 
\vspace{1mm}
$\;\;\;\;\;=|\underbrace{\sgenrowbox}_{2j_{L}}, 
\underbrace{\sgenrowbox}_{2j_{R}} 
\rangle.$\\
\vspace{1mm}\\
The corresponding supermultiplets are displayed in table 12 of 
ref \cite{gmz1}. Again they involve spins and AdS energies that increase 
with $j_{L}$ and $j_{R}$, maintaining a constant spin range of 4  for 
$j_{L}, j_{R}\geq 1$.

\end{itemize}
In addition to these purely (super)symmetrized lwv's, one can also 
anti-(super)symmetrize pairs of superoscillators, since $P=2$. The requirement 
$L^{-}|\Omega\rangle =0$
then rules out the simultaneous appearance of $\xi$'s and $\eta$'s so that 
one is left with

\begin{itemize}
\item $|\Omega\rangle = 
\xi^{[A_{1}}(1)\xi^{B_{1}]}(2)...\xi^{[A_{n}}(1) \xi^{B_{n}]}(2) 
\xi^{C_{1}}(1)\dots \xi^{C_{k}}(1) |0\rangle$\\
\vspace{1mm} 
$\;\;\;\;\; =|\marcshapirowbox,1\rangle$
\vspace{2mm}
\item $|\Omega\rangle = 
\eta^{[A_{1}}(1)\eta^{B_{1}]}(2)...\eta^{[A_{n}}(1) \eta^{B_{n}]}(2) 
\eta^{C_{1}}(1)\dots \eta^{C_{k}}(1) |0\rangle$\\ 
\vspace{1mm}
$\;\;\;\;\;=|1,\marcshapirowbox\rangle$.

\end{itemize}
\vspace{1.5mm}
The special case $k=0$ then leads to the novel short multiplets, which we  
will now discuss in detail.

The simplest case is
\begin{eqnarray}
|\Omega \rangle& = & \xi^{[A_{1}}(1)\xi^{B_{1}]}(2)|0\rangle= |\soneonebox, 1
\rangle\nonumber\\
               &\equiv & a^{[i}(1)a^{j]}(2)|0\rangle\oplus[a^{i}(1)
{\alpha}^{\gamma}(2)-a^{i}(2){\alpha}^{\gamma}(1)]|0\rangle \oplus 
{\alpha}^{(\gamma}
(1){\alpha}^{\delta)}(2)|0\rangle.\nonumber
\end{eqnarray}
Acting on $|\Omega\rangle$ with the supersymmetry generators 
${\vec{a}}^{i}\cdot 
{\vec{\beta}}^{x}$ and ${\vec{b}}^{r}\cdot{\vec{\alpha}}^{\gamma}$ of 
$L^{+}$ and collecting resulting $SU(2,2)\times SU(4)$ lwv's 
(i.e. states that are 
annihilated by ${\vec{a}}_{i}\cdot 
{\vec{b}}_{r}$ and ${\vec{\alpha}}_{\gamma}\cdot 
{\vec{\beta}}_{x}$), one arrives at the supermultiplet of spin range 2 given 
in Table 6 (see the Appendix for a complete list of the allowed $SU(4)$
lwv's for $P=2$ and the corresponding $SU(4)$ representations they induce).

\vspace{.2cm}
\begin{center}
\begin{tabular}{|c|c|c|c|c|}
\hline
E =-l                & ($j_{L},j_{R}$)=($j_M,j_N$)     & SU(4) Dynkin     & 
$Y$
& Field\\ \hline
2               & (0,0)               & (2,0,0)         & 2 &$\Phi_{0,0}$
\\ \hline
5/2           & (0,1/2)           & (1,0,0) & 3 &$\Psi_{0,1/2}$
\\ \hline
5/2           & (1/2,0)           & (1,1,0)         & 1 &$\Psi_{1/2,0}$
\\ \hline
3             & (1 ,0)            & (1,0,1) & 0 &$\Phi_{1,0}$
\\ \hline
3                & (1/2,1/2)             & (0,1,0)        & 2 &$\Phi_{1/2,1/2}$
\\ \hline
3               & (0,0)               & (0,2,0)         & 0 &$\Phi_{0,0}$
\\ \hline
3           & (0,0)           & (0,0,0) & 4 &$\Phi_{0,0}$
\\ \hline
7/2           & (1/2,0)           & (0,1,1)         & -1 &$\Psi_{1/2,0}$
\\ \hline
7/2             & (3/2 ,0)            & (1,0,0) & -1 &$\Psi_{3/2,0}$
\\ \hline
7/2                & (1,1/2)             & (0,0,1)     & 1 &$\Psi_{1,1/2}$
\\ \hline
4               & (1,0)               & (0,1,0)         & -2 &$\Phi_{1,0}$
\\ \hline
4          & (0,0)           & (0,0,2) & -2 &$\Phi_{0,0}$
\\ \hline
4          & (3/2,1/2)           & (0,0,0) & 0 &$\Phi_{3/2,1/2}$
\\ \hline
9/2           & (1/2,0)           & (0,0,1)         & -3 &$\Psi_{1/2,0}$
\\ \hline
5             & (0,0)            & (0,0,0) & -4 &$\Phi_{0,0}$
\\ \hline
\end{tabular}
\end{center}
Table 6. The supermultiplet
corresponding to the lwv 
$|\Omega \rangle = \xi^{[A_{1}}(1)\xi^{B_{1}]}(2)|0\rangle $.
It has $Z=1$.
\vspace{.7cm}

This supermultiplet does not occur in the tensor product of two or more 
CPT self-conjugate doubleton supermultiplets, but it appears in the tensor 
product of two doubleton supermultiplets of the type listed in table 2.

Similarly, by taking the following lwv
\begin{eqnarray}
|\Omega \rangle &= &\eta^{[A_{1}}(1)\eta^{B_{1}]}(2)|0\rangle=
|1,\soneonebox\rangle\nonumber\\
&\equiv&b^{[r}(1)b^{s]}(2)|0\rangle\oplus[b^{r}(1)
{\beta}^{x}(2)-b^{r}(2){\beta}^{x}(1)]|0\rangle \oplus {\beta}^{(x}
(1){\beta}^{y)}(2)|0\rangle,\nonumber
\end{eqnarray}
one gets the CPT conjugate supermultiplet  displayed in Table 7.

\vspace{.2cm}
\begin{center}
\begin{tabular}{|c|c|c|c|c|}
\hline
E=-l                & ($j_{L},j_{R}$)=($j_M,j_N$)    & SU(4) Dynkin     
& $Y$
& Field\\ \hline
2               & (0,0)               & (0,0,2)         &-2 &$\Phi_{0,0}$
\\ \hline
5/2           & (1/2,0)           & (0,0,1) &-3 &$\Psi_{1/2,0}$
\\ \hline
5/2           & (0,1/2)           & (0,1,1)         & -1 &$\Psi_{0,1/2}$
\\ \hline
3             & (0,1)            & (1,0,1) & 0 &$\Phi_{0,1}$
\\ \hline
3                & (1/2,1/2)             & (0,1,0)        &-2 &$\Phi_{1/2,1/2}$
\\ \hline
3               & (0,0)               & (0,2,0)         & 0 &$\Phi_{0,0}$
\\ \hline
3           & (0,0)           & (0,0,0) & -4 &$\Phi_{0,0}$
\\ \hline
7/2           & (0,1/2)           & (1,1,0)         & 1 &$\Psi_{0,1/2}$
\\ \hline
7/2             & (0,3/2)            & (0,0,1) & 1 &$\Psi_{0,3/2}$
\\ \hline
7/2                & (1/2,1)             & (1,0,0)     & -1 &$\Psi_{1/2,1}$
\\ \hline
4               & (0,1)               & (0,1,0)         & 2 &$\Phi_{0,1}$
\\ \hline
4          & (0,0)           & (2,0,0) & 2 &$\Phi_{0,0}$
\\ \hline
4          & (1/2,3/2)           & (0,0,0) & 0 &$\Phi_{1/2,3/2}$
\\ \hline
9/2           & (0,1/2)           & (1,0,0)         & 3 &$\Psi_{0,1/2}$
\\ \hline
5             & (0,0)            & (0,0,0) & 4 &$\Phi_{0,0}$
\\ \hline
\end{tabular}
\end{center}
Table 7. The supermultiplet
corresponding to the lwv 
$|\Omega \rangle = \eta^{[A_{1}}(1)\eta^{B_{1}]}(2)|0\rangle $.
It has $Z=-1$.
\vspace{.7cm}

In complete analogy to its CPT conjugate counterpart, it occurs in the 
tensor product of two doubleton supermultiplets of the type given in table 3,
but not in the tensor product of two or more self-conjugate doubleton 
supermultiplets.

The general lwv for $j \geq 2$  
\eqn
|\Omega\rangle &=& 
\xi^{[A_{1}}(1)\xi^{B_{1}]}(2)...\xi^{[A_{j}}(1) \xi^{B_{j}]}(2) 
|0\rangle\nn\cr
&=& |\underbrace{\sgenrowrowbox}_{j}, 1 \rangle
\enn
leads to the following supermultiplet with spin range 2

\vspace{.2cm}
\begin{center}
\begin{tabular}{|c|c|c|c|c|}
\hline
E =-l                & ($j_{L},j_{R}$)=($j_M,j_N$)     & SU(4) Dynkin    
& $Y$
& Field\\ \hline
j               & (0,0)               & (0,0,0)         & 4 &$\Phi_{0,0}$
\\ \hline
j+1/2           & (1/2,0)           & (1,0,0)         & 3 &$\Psi_{1/2,0}$
\\ \hline
j+1               & (0,0)               & (2,0,0)         & 2 &$\Phi_{0,0}$
\\ \hline
j+1             & (1 ,0)            & (0,1,0) & 2 &$\Phi_{1,0}$
\\ \hline
j+3/2           & (1/2,0)           & (1,1,0)         & 1 &$\Psi_{1/2,0}$
\\ \hline
j+3/2             & (3/2 ,0)            & (0,0,1) & 1 &$\Psi_{3/2,0}$
\\ \hline
j+2               & (2,0)               & (0,0,0)         & 0 &$\Phi_{2,0}$
\\ \hline
j+2               & (1,0)               & (1,0,1)         & 0 &$\Phi_{1,0}$
\\ \hline
j+2          & (0,0)           & (0,2,0) & 0 &$\Phi_{0,0}$
\\ \hline
j+5/2           & (3/2,0)           & (1,0,0)         & -1 &$\Psi_{3/2,0}$
\\ \hline
j+5/2           & (1/2,0)           & (0,1,1)         & -1 &$\Psi_{1/2,0}$
\\ \hline
j+3           & (1,0)           & (0,1,0)         & -2 &$\Phi_{1,0}$
\\ \hline
j+3           & (0,0)           & (0,0,2)         & -2 &$\Phi_{0,0}$
\\ \hline
j+7/2             & (1/2,0)            & (0,0,1) & -3 &$\Psi_{1/2,0}$
\\ \hline
j+4             & (0,0)            & (0,0,0) & -4 &$\Phi_{0,0}$
\\ \hline
\end{tabular}
\end{center}
Table 8. The supermultiplet
corresponding to the lwv
 
$|\Omega \rangle =
\xi^{[A_{1}}(1)\xi^{B_{1}]}(2)...\xi^{[A_{j}}(1) \xi^{B_{j}]}(2) |0\rangle  $.
It has $Z=j$.
\vspace{.7cm}

Obviously, the spin  content of these multiplets is independent of 
$j$. Only the AdS energies (resp. conformal dimensions) get shifted, when 
$j$ is increased, which distinguishes these multiplets from their 
(super)symmetrized counterparts obtained from $|\Omega \rangle = 
\xi^{A_1}(1) \xi^{A_2}(1) ... \xi^{A_{2j}}(1) |0\rangle$, where the spins 
increase with $j$.

Similarly, the conjugate lwv ($j\geq 2$)
\eqn
|\Omega\rangle &= &
\eta^{[A_{1}}(1)\eta^{B_{1}]}(2)...\eta^{[A_{j}}(1) \eta^{B_{j}]}(2) 
|0\rangle\nn\cr
& = & |1, \underbrace{\sgenrowrowbox}_{j}\rangle
\enn
leads to the supermultiplet given in Table 9 

\vspace{.2cm}
\begin{center}
\begin{tabular}{|c|c|c|c|c|}
\hline
E=-l                 & ($j_{L},j_{R}$)=($j_M,j_N$)     & SU(4) Dynkin    
& $Y$
& Field\\ \hline
j               & (0,0)               & (0,0,0)         & -4 &$\Phi_{0,0}$
\\ \hline
j+1/2           & (0,1/2)           & (0,0,1)         & -3 &$\Psi_{0,1/2}$
\\ \hline
j+1               & (0,0)               & (0,0,2)         & -2 &$\Phi_{0,0}$
\\ \hline
j+1             & (0,1)            & (0,1,0) & -2 &$\Phi_{0,1}$
\\ \hline
j+3/2           & (0,1/2)           & (0,1,1)         & -1 &$\Psi_{0,1/2}$
\\ \hline
j+3/2             & (0,3/2)            & (1,0,0) & -1 &$\Psi_{0,3/2}$
\\ \hline
j+2               & (0,2)               & (0,0,0)         & 0 &$\Phi_{0,2}$
\\ \hline
j+2               & (0,1)               & (1,0,1)         & 0 &$\Phi_{0,1}$
\\ \hline
j+2          & (0,0)           & (0,2,0) & 0 &$\Phi_{0,0}$
\\ \hline
j+5/2           & (0,3/2)           & (0,0,1)         & 1 &$\Psi_{0,3/2}$
\\ \hline
j+5/2           & (0,1/2)           & (1,1,0)         & 1 &$\Psi_{0,1/2}$
\\ \hline
j+3           & (0,1)           & (0,1,0)         & 2 &$\Phi_{0,1}$
\\ \hline
j+3           & (0,0)           & (2,0,0)         & 2 &$\Phi_{0,0}$
\\ \hline
j+7/2             & (0,1/2)            & (1,0,0) & 3 &$\Psi_{0,1/2}$
\\ \hline
j+4             & (0,0)            & (0,0,0) & 4 &$\Phi_{0,0}$
\\ \hline
\end{tabular}
\end{center}
Table 9. The supermultiplet
corresponding to the lwv 

$|\Omega \rangle = \eta^{[A_{1}}(1)\eta^{B_{1}]}(2)
...\eta^{[A_{j}}(1)\eta^{B_{j}]}(2) |0\rangle $.
It has $Z=-j$.
\vspace{.7cm}

\section{Discussion and Conclusions}
\setcounter{equation}{0}
The spectrum of the $S^5$ compactification of  IIB supergravity  falls 
into an infinite tower of irreducible CPT self-conjugate 
supermultiplets of $SU(2,2|4)$ of spin range two with ever increasing 
quantized eigenvalues of AdS energy \cite{mgnm,krv} . They are obtained
by choosing as lowest weight vector the (super) Fock vacuum vector 
$|0\rangle$ of ever increasing pairs $P\geq 2$ of super-oscillators 
and therefore have $Z=0$. Hence they are representations of $PSU(2,2|4)$.
The ``massless'' 
graviton supermultiplet of ${\cal{N}}= 8$ AdS supergravity in $d=5$ sits at 
the 
bottom of this infinite tower corresponding to $P=2$.  The higher AdS energy supermultiplets 
correspond to ``massive'' Kaluza-Klein modes. The shortest irreducible CPT 
self-conjugate supermultiplet of spin range one ($P=1$) decouples from the 
Kaluza-Klein spectrum as local gauge degrees of freedom. This 
ultrashort supermultiplet is the unique CPT self-conjugate irreducible 
doubleton 
supermultiplet, which has no smooth Poincar\'{e} limit in $d=5$. It lives on 
the boundary of $AdS_5$, which can be identified with  four 
dimensional Minkowski space. 

 The quadratic operator that reduces to 
the mass (squared) operator in $d=5$ Minkowski space when one takes the 
Poincar\'{e} limit of $AdS_{5}$ is not a Casimir invariant of $SU(2,2)$. 
Hence the 
Poincar\'{e} mass is not an invariant quantity in $AdS_5$. One may, 
instead,  use the eigenvalues of the quadratic, cubic and quartic  
Casimir operators as  invariant labels of a UIR of $SU(2,2)$. However, 
for positive energy UIR's one can use the labels ($j_L,j_R, E$) of the 
corresponding lowest weight vectors $|\Omega \rangle $ with respect to 
the maximal compact subgroup $SU(2)_L \times SU(2)_R \times U(1)_{E}$ . We 
showed in section 2 explicitly  how one can go to a non-compact basis 
for the UIR defined by the lowest weight vector such that the UIR is 
now labelled with respect to the $SL(2,\mathbf{C}) \times \mathcal{D}$ 
subgroup with labels ($j_M,j_N,l$), which numerically coincide with 
$(j_L,j_R,-E)$. 
For doubleton fields living on the boundary, the group $SU(2,2)$ acts as 
the conformal group and hence the labels  ($j_M,j_N$)  are the 
covariant
labels with respect to the Lorentz group in $d=4$ and $l=-E$ is simply 
the scaling (conformal) dimension. We verified that the doubleton 
irreps correspond to massless fields in $d=4$. Non-doubleton 
representations
($P >1$) are the massive representations of the four dimensional 
conformal group. Interpreted as the anti-de Sitter group in $d=5$ some 
of these representations of $SU(2,2)$ with $P=2$ become massless 
representations in the ($5d$) Poincar\'{e} limit. These are precisely the 
representations that satisfy the condition $E=j_L+j_R+2$.

In this paper we focussed mainly on some novel short supermultiplets of 
$SU(2,2|4)$ that are not CPT self-conjugate. These supermultiplets cannot 
be obtained by tensoring  CPT self-conjugate ${\cal{N}}=4$ Yang-Mills 
doubleton supermultiplets with themselves. For $P=2$ these short 
supermultiplets involve fields that do not satisfy the condition 
$E=j_L+j_R+2$.
Interestingly, all these novel short supermultiplets of the form we 
discussed
in section 5 can be obtained by tensoring the  chiral doubleton 
supermultiplets given in Tables 2 to 5 with themselves. One may ask what role,
if any,  these novel supermultiplets play in the AdS/CFT duality. As 
we argued above, we expect the 
massive supermultiplets of $1/4$ BPS states in ${\cal{N}}=4$ super Yang-Mills
theory to decompose into a CPT conjugate pair of chiral spin 3/2 
doubleton supermultiplets plus two CPT self-conjugate doubleton 
supermultiplets of $SU(2,2|4)$, when an appropriate conformal limit
(given its existence) is taken. Now, the $1/4$  BPS states belong
to nontrivial orbits of the duality group $SL(2, \mathbf{Z})$ and
are related to $(p,q)$ IIB superstrings \cite{pqstring}. This suggests that 
the novel supermultiplets we studied above for $P=2$ as well 
as their counterparts for $P>2$ are relevant to a generalized duality
between the solitonic sector of ${\cal{N}}=4$ super Yang-Mills in $d=4$
and the $(p,q)$ superstrings over $AdS_5 \times S^5$.   More generally,
the methods  we have employed in this paper provide us with  simple and yet
powerful tools for answering the question whether it is possible to extend
the $AdS/CFT$ duality to the full $SL(2, \mathbf{Z})$ covariant type IIB
superstring over $AdS_5 \times S^5$.

{\bf Acknowledgements}
We would like to thank Sergio Ferrara and Juan Maldacena
for enlightening  discussions.

\section{Appendix}
\setcounter{equation}{0}

The allowed lowest weight vectors (lwv's) of $SU(4)$ for $P=2$ (i.e. the 
states annihilated by $L^{-}_{\gamma x}= {\vec{\alpha}}_{\gamma} 
\cdot {\vec{\beta}}_{x}$) are given in the
following table. This table is more explicit than the corresponding table 
in the Appendix of our previous paper \cite{gmz1} in that it now also contains
all non-trivial realizations in terms of some less obvious linear 
combinations (like the one in the last row, for example). 
These linear combinations do not lead to any new $SU(4)$ representation,
however they can sometimes conspire to give some non-trivially realized 
states in supermultiplets. In our previous paper \cite{gmz1} this was the 
case for the multiplet listed in Table 12. In the present paper, 
on the other hand, 
they do not contribute
due to the (super)antisymmetry of the lwv's considered in this paper: In the  
cases where they might contribute, one always encounters a wrong minus sign.

\vspace{.2cm}
\begin{center}
\begin{tabular}{|c|c|c|}
\hline
\multicolumn{1}{|c|}{lwv} &\multicolumn{1}{|c|}{SU(4) Dynkin (dim)}   
&\multicolumn{1}{|c|}{$Y$}
\\ \hline
$|0\rangle $ &(0,2,0) (20') & 0
\\ \hline
$\beta^x(1) |0\rangle $  &(0,1,1) ($\bar{20}$)  & -1
\\ \hline
$\alpha^{\gamma}(1) |0\rangle$ & (1,1,0) (20)   & 1
\\ \hline
$\beta^x (1) \beta^y (1) |0\rangle $ &(0,1,0) (6) & -2
\\ \hline
$\alpha^{\gamma} (1) \alpha^{\delta} (1) |0\rangle $ &(0,1,0) (6)  & 2
\\ \hline
$\beta^x \beta^y \beta^z \beta^w |0\rangle$  &(0,0,0) (1) &-4
\\ \hline
$\alpha^{\gamma} \alpha^{\delta} \alpha^{\epsilon} \alpha^{\eta}|0\rangle $
& (0,0,0) (1) &4
\\ \hline
$\beta^{(x}(1) \beta^{y)}(2) |0\rangle$  &(0,0,2) ($\bar{10}$) & -2
\\ \hline
$\alpha^{(\gamma}(1) \alpha^{\delta)}(2)|0\rangle$ &(2,0,0) (10) & 2
\\ \hline
$\beta^x \beta^y \beta^z |0\rangle$ &(0,0,1) ($\bar{4}$) & -3
\\ \hline
$\alpha^{\gamma} \alpha^{\delta} \alpha^{\epsilon} |0\rangle$ &(1,0,0) (4) &3
\\ \hline
$\alpha^{\gamma}(1)\beta^x(2) |0\rangle$ &(1,0,1) (15) &0
\\ \hline
$[\alpha^{\gamma}(1)\beta^x(1)
- \alpha^{\gamma}(2)\beta^x(2)] |0\rangle$ &(1,0,1) (15) &0
\\ \hline
$\alpha^{\gamma}(1)\beta^x(2) \beta^y(2)|0\rangle$ &(1,0,0) (4) &-1
\\ \hline
$[\alpha^{\gamma}(1)\beta^{[x}(1) \beta^{y]}(2)$ 
$- {1 \over 2} \alpha^{\gamma}(2)\beta^x(2) \beta^y(2)]|0\rangle$ 
& (1,0,0) (4) & -1
\\ \hline
$\beta^x(1)\alpha^{\gamma}(2) \alpha^{\delta}(2) |0\rangle$ &(0,0,1) 
($\bar{4}$) &1
\\ \hline
$[\beta^x(1)\alpha^{[ \gamma}(1) \alpha^{\delta ]}(2)$  
& (0,0,1) ($\bar{4}$) & 1 \\
$- {1 \over 2} \beta^x(2)\alpha^{\gamma}(2) \alpha^{\delta}(2) ]|0\rangle$ 
&&
\\ \hline
$\alpha^{\gamma}(1) \alpha^{\delta}(1) \beta^x(2) \beta^y(2) |0\rangle$
&(0,0,0) (1) &0
\\ \hline
$[\alpha^{\gamma}(1) \alpha^{\delta}(1) \beta^{[x}(1) \beta^{y]}(2)$ 
& (0,0,0) (1) & 0 \\ 
$- \alpha^{[ \gamma}(2) \alpha^{\delta ]}(1) \beta^{x}(2) 
\beta^{y}(2)]|0\rangle$ & & \\ \hline
$[\alpha^{[ \gamma}(1) \alpha^{\delta ]}(2) \beta^{[x}(1) \beta^{y]}(2)$ 
& (0,0,0)  (1) &  0 \\  
$ - {1 \over 4} \alpha^{\gamma}(1) \alpha^{\delta}(1) \beta^x(1) 
\beta^y(1)$ & & \\
$- {1 \over 4} \alpha^{\gamma}(2) \alpha^{\delta}(2) \beta^x(2) 
\beta^y(2)]|0\rangle$ & &  \\ \hline
 
\end{tabular}
\end{center}
\vspace{.2cm}

For the decomposition of the supertableaux of $U(m/n)$ in terms of the
tableaux of its even subgroup $U(m)\times U(n)$ we refer to \cite{bbars}. Here
we give a few examples.

\begin{equation}
\begin{array}{rl}
~&~\\
U(m/n) & \supset U(m)\times U(n)\\
~&~\\
\sonebox & = (\onebox,1)+(1,\onebox)\\
~&~\\
\stwobox & = (\twobox,1)+(\onebox,\onebox)+(1,\oneonebox)\\
~&~\\
\soneonebox & = (\oneonebox,1)+(\onebox,\onebox)+(1,\twobox)\\
~&~\\
\stwoonebox & = (\twoonebox,1)+(\twobox,\onebox)+(\onebox,\oneonebox)\\
~&~\\
~& +(1,\twoonebox)+(\oneonebox,\onebox)+(\onebox,\twobox)\\
~&~
\end{array}
\end{equation}

\end{document}